\documentclass[aps,prd,twocolumn,amsmath,amssymb,amsfonts,nofootinbib,superscriptaddress,altaffilletter]{revtex4-1}
\usepackage{graphicx}
\usepackage{hyperref}
\hypersetup{
  bookmarksopen=true
}
\usepackage{amssymb}
\usepackage{amsmath}
\usepackage{color}
\usepackage{supertabular}
\usepackage{dcolumn}

\def\Tobs{T_{\textrm{\mbox{\tiny{obs}}}}}
\def\Tcoh{T_{\textrm{\mbox{\tiny{coh}}}}}

\def\EatH{Einstein@Home }

\def\ec{\textrm{~,}}

\def\fdot{f^{(1)}}

\def\beq{\begin{equation}}
\def\eeq{\end{equation}}
\def\bea{\begin{eqnarray}}
\def\eea{\end{eqnarray}}

\newcommand{\F}{\mathcal{F}}
\newcommand{\SNR}{\textrm{SNR}}

\newcommand{\Fstat}{$\mathcal{F}$-statistic}
\def\td{{\it Time-Domain \Fstat}}

\def\sci#1#2{#1\times10^{#2}}

\def\RAJ{\textrm{RA}_{\textrm J2000}}
\def\DECJ{\textrm{DEC}_{\textrm J2000}}

\begin{document}

\title{ 
Full Band All-sky Search for Periodic Gravitational Waves in the O1 LIGO Data.
}




\author{%
B.~P.~Abbott,$^{1}$  
R.~Abbott,$^{1}$  
T.~D.~Abbott,$^{2}$  
F.~Acernese,$^{3,4}$ 
K.~Ackley,$^{5,6}$  
C.~Adams,$^{7}$  
T.~Adams,$^{8}$ 
P.~Addesso,$^{9}$  
R.~X.~Adhikari,$^{1}$  
V.~B.~Adya,$^{10}$
% ...
\\
\medskip
(LIGO Scientific Collaboration and Virgo Collaboration) 
\\
\medskip
{${}^{\ast}$Deceased, February 2017. }%
{${}^{\dag}$Deceased, November 2017. }%
{${}^{\ddag}$Deceased, December 2016. }%
}\noaffiliation
\affiliation {LIGO, California Institute of Technology, Pasadena, CA 91125, USA }
\affiliation {Louisiana State University, Baton Rouge, LA 70803, USA }
\affiliation {Universit\`a di Salerno, Fisciano, I-84084 Salerno, Italy }
\affiliation {INFN, Sezione di Napoli, Complesso Universitario di Monte S.Angelo, I-80126 Napoli, Italy }
\affiliation {University of Florida, Gainesville, FL 32611, USA }
\affiliation {OzGrav, School of Physics \& Astronomy, Monash University, Clayton 3800, Victoria, Australia }
\affiliation {LIGO Livingston Observatory, Livingston, LA 70754, USA }
\affiliation {Laboratoire d'Annecy-le-Vieux de Physique des Particules (LAPP), Universit\'e Savoie Mont Blanc, CNRS/IN2P3, F-74941 Annecy, France }
\affiliation {University of Sannio at Benevento, I-82100 Benevento, Italy and INFN, Sezione di Napoli, I-80100 Napoli, Italy }
\affiliation {Max Planck Institute for Gravitational Physics (Albert Einstein Institute), D-30167 Hannover, Germany }
\affiliation {The University of Mississippi, University, MS 38677, USA }
\affiliation {NCSA, University of Illinois at Urbana-Champaign, Urbana, IL 61801, USA }
\affiliation {University of Cambridge, Cambridge CB2 1TN, United Kingdom }
\affiliation {Nikhef, Science Park, 1098 XG Amsterdam, The Netherlands }
\affiliation {LIGO, Massachusetts Institute of Technology, Cambridge, MA 02139, USA }
\affiliation {Instituto Nacional de Pesquisas Espaciais, 12227-010 S\~{a}o Jos\'{e} dos Campos, S\~{a}o Paulo, Brazil }
\affiliation {Gran Sasso Science Institute (GSSI), I-67100 L'Aquila, Italy }
\affiliation {INFN, Laboratori Nazionali del Gran Sasso, I-67100 Assergi, Italy }
\affiliation {Inter-University Centre for Astronomy and Astrophysics, Pune 411007, India }
\affiliation {University of Wisconsin-Milwaukee, Milwaukee, WI 53201, USA }
\affiliation {Leibniz Universit\"at Hannover, D-30167 Hannover, Germany }
\affiliation {Universit\`a di Pisa, I-56127 Pisa, Italy }
\affiliation {INFN, Sezione di Pisa, I-56127 Pisa, Italy }
\affiliation {OzGrav, Australian National University, Canberra, Australian Capital Territory 0200, Australia }
\affiliation {Laboratoire des Mat\'eriaux Avanc\'es (LMA), CNRS/IN2P3, F-69622 Villeurbanne, France }
\affiliation {SUPA, University of the West of Scotland, Paisley PA1 2BE, United Kingdom }
\affiliation {LAL, Univ. Paris-Sud, CNRS/IN2P3, Universit\'e Paris-Saclay, F-91898 Orsay, France }
\affiliation {California State University Fullerton, Fullerton, CA 92831, USA }
\affiliation {European Gravitational Observatory (EGO), I-56021 Cascina, Pisa, Italy }
\affiliation {Universit\`a di Roma Tor Vergata, I-00133 Roma, Italy }
\affiliation {INFN, Sezione di Roma Tor Vergata, I-00133 Roma, Italy }
\affiliation {Universit\"at Hamburg, D-22761 Hamburg, Germany }
\affiliation {INFN, Sezione di Roma, I-00185 Roma, Italy }
\affiliation {Cardiff University, Cardiff CF24 3AA, United Kingdom }
\affiliation {Embry-Riddle Aeronautical University, Prescott, AZ 86301, USA }
\affiliation {Max Planck Institute for Gravitational Physics (Albert Einstein Institute), D-14476 Potsdam-Golm, Germany }
\affiliation {APC, AstroParticule et Cosmologie, Universit\'e Paris Diderot, CNRS/IN2P3, CEA/Irfu, Observatoire de Paris, Sorbonne Paris Cit\'e, F-75205 Paris Cedex 13, France }
\affiliation {Korea Institute of Science and Technology Information, Daejeon 34141, Korea }
\affiliation {West Virginia University, Morgantown, WV 26506, USA }
\affiliation {Universit\`a di Perugia, I-06123 Perugia, Italy }
\affiliation {INFN, Sezione di Perugia, I-06123 Perugia, Italy }
\affiliation {Syracuse University, Syracuse, NY 13244, USA }
\affiliation {University of Minnesota, Minneapolis, MN 55455, USA }
\affiliation {SUPA, University of Glasgow, Glasgow G12 8QQ, United Kingdom }
\affiliation {LIGO Hanford Observatory, Richland, WA 99352, USA }
\affiliation {Caltech CaRT, Pasadena, CA 91125, USA }
\affiliation {Wigner RCP, RMKI, H-1121 Budapest, Konkoly Thege Mikl\'os \'ut 29-33, Hungary }
\affiliation {Columbia University, New York, NY 10027, USA }
\affiliation {Stanford University, Stanford, CA 94305, USA }
\affiliation {Universit\`a di Camerino, Dipartimento di Fisica, I-62032 Camerino, Italy }
\affiliation {Universit\`a di Padova, Dipartimento di Fisica e Astronomia, I-35131 Padova, Italy }
\affiliation {INFN, Sezione di Padova, I-35131 Padova, Italy }
\affiliation {Institute of Physics, E\"otv\"os University, P\'azm\'any P. s. 1/A, Budapest 1117, Hungary }
\affiliation {Nicolaus Copernicus Astronomical Center, Polish Academy of Sciences, 00-716, Warsaw, Poland }
\affiliation {Rochester Institute of Technology, Rochester, NY 14623, USA }
\affiliation {University of Birmingham, Birmingham B15 2TT, United Kingdom }
\affiliation {INFN, Sezione di Genova, I-16146 Genova, Italy }
\affiliation {RRCAT, Indore MP 452013, India }
\affiliation {Faculty of Physics, Lomonosov Moscow State University, Moscow 119991, Russia }
\affiliation {SUPA, University of Strathclyde, Glasgow G1 1XQ, United Kingdom }
\affiliation {The Pennsylvania State University, University Park, PA 16802, USA }
\affiliation {OzGrav, University of Western Australia, Crawley, Western Australia 6009, Australia }
\affiliation {Department of Astrophysics/IMAPP, Radboud University Nijmegen, P.O. Box 9010, 6500 GL Nijmegen, The Netherlands }
\affiliation {Artemis, Universit\'e C\^ote d'Azur, Observatoire C\^ote d'Azur, CNRS, CS 34229, F-06304 Nice Cedex 4, France }
\affiliation {Institut FOTON, CNRS, Universit\'e de Rennes 1, F-35042 Rennes, France }
\affiliation {Washington State University, Pullman, WA 99164, USA }
\affiliation {University of Oregon, Eugene, OR 97403, USA }
\affiliation {Laboratoire Kastler Brossel, UPMC-Sorbonne Universit\'es, CNRS, ENS-PSL Research University, Coll\`ege de France, F-75005 Paris, France }
\affiliation {Carleton College, Northfield, MN 55057, USA }
\affiliation {OzGrav, University of Adelaide, Adelaide, South Australia 5005, Australia }
\affiliation {Astronomical Observatory Warsaw University, 00-478 Warsaw, Poland }
\affiliation {VU University Amsterdam, 1081 HV Amsterdam, The Netherlands }
\affiliation {University of Maryland, College Park, MD 20742, USA }
\affiliation {School of Physics, Georgia Institute of Technology, Atlanta, GA 30332, USA }
\affiliation {Universit\'e Claude Bernard Lyon 1, F-69622 Villeurbanne, France }
\affiliation {Universit\`a di Napoli `Federico II,' Complesso Universitario di Monte S.Angelo, I-80126 Napoli, Italy }
\affiliation {NASA Goddard Space Flight Center, Greenbelt, MD 20771, USA }
\affiliation {Dipartimento di Fisica, Universit\`a degli Studi di Genova, I-16146 Genova, Italy }
\affiliation {RESCEU, University of Tokyo, Tokyo, 113-0033, Japan. }
\affiliation {Tsinghua University, Beijing 100084, China }
\affiliation {Texas Tech University, Lubbock, TX 79409, USA }
\affiliation {Kenyon College, Gambier, OH 43022, USA }
\affiliation {Departamento de Astronom\'{\i }a y Astrof\'{\i }sica, Universitat de Val\`encia, E-46100 Burjassot, Val\`encia, Spain }
\affiliation {Museo Storico della Fisica e Centro Studi e Ricerche Enrico Fermi, I-00184 Roma, Italy }
\affiliation {National Tsing Hua University, Hsinchu City, 30013 Taiwan, Republic of China }
\affiliation {Charles Sturt University, Wagga Wagga, New South Wales 2678, Australia }
\affiliation {Center for Interdisciplinary Exploration \& Research in Astrophysics (CIERA), Northwestern University, Evanston, IL 60208, USA }
\affiliation {University of Chicago, Chicago, IL 60637, USA }
\affiliation {Pusan National University, Busan 46241, Korea }
\affiliation {The Chinese University of Hong Kong, Shatin, NT, Hong Kong }
\affiliation {INAF, Osservatorio Astronomico di Padova, I-35122 Padova, Italy }
\affiliation {INFN, Trento Institute for Fundamental Physics and Applications, I-38123 Povo, Trento, Italy }
\affiliation {OzGrav, University of Melbourne, Parkville, Victoria 3010, Australia }
\affiliation {Universit\`a di Roma `La Sapienza,' I-00185 Roma, Italy }
\affiliation {Universit\'e Libre de Bruxelles, Brussels 1050, Belgium }
\affiliation {Sonoma State University, Rohnert Park, CA 94928, USA }
\affiliation {Departamento de Matem\'aticas, Universitat de Val\`encia, E-46100 Burjassot, Val\`encia, Spain }
\affiliation {Montana State University, Bozeman, MT 59717, USA }
\affiliation {Universitat de les Illes Balears, IAC3---IEEC, E-07122 Palma de Mallorca, Spain }
\affiliation {The University of Texas Rio Grande Valley, Brownsville, TX 78520, USA }
\affiliation {Bellevue College, Bellevue, WA 98007, USA }
\affiliation {Institute for Plasma Research, Bhat, Gandhinagar 382428, India }
\affiliation {The University of Sheffield, Sheffield S10 2TN, United Kingdom }
\affiliation {Dipartimento di Scienze Matematiche, Fisiche e Informatiche, Universit\`a di Parma, I-43124 Parma, Italy }
\affiliation {INFN, Sezione di Milano Bicocca, Gruppo Collegato di Parma, I-43124 Parma, Italy }
\affiliation {California State University, Los Angeles, 5151 State University Dr, Los Angeles, CA 90032, USA }
\affiliation {Universit\`a di Trento, Dipartimento di Fisica, I-38123 Povo, Trento, Italy }
\affiliation {NCBJ, 05-400 \'Swierk-Otwock, Poland }
\affiliation {Montclair State University, Montclair, NJ 07043, USA }
\affiliation {National Astronomical Observatory of Japan, 2-21-1 Osawa, Mitaka, Tokyo 181-8588, Japan }
\affiliation {Canadian Institute for Theoretical Astrophysics, University of Toronto, Toronto, Ontario M5S 3H8, Canada }
\affiliation {Observatori Astron\`omic, Universitat de Val\`encia, E-46980 Paterna, Val\`encia, Spain }
\affiliation {School of Mathematics, University of Edinburgh, Edinburgh EH9 3FD, United Kingdom }
\affiliation {University and Institute of Advanced Research, Koba Institutional Area, Gandhinagar Gujarat 382007, India }
\affiliation {IISER-TVM, CET Campus, Trivandrum Kerala 695016, India }
\affiliation {University of Szeged, D\'om t\'er 9, Szeged 6720, Hungary }
\affiliation {International Centre for Theoretical Sciences, Tata Institute of Fundamental Research, Bengaluru 560089, India }
\affiliation {University of Michigan, Ann Arbor, MI 48109, USA }
\affiliation {Tata Institute of Fundamental Research, Mumbai 400005, India }
\affiliation {INAF, Osservatorio Astronomico di Capodimonte, I-80131, Napoli, Italy }
\affiliation {Universit\`a degli Studi di Urbino `Carlo Bo,' I-61029 Urbino, Italy }
\affiliation {INFN, Sezione di Firenze, I-50019 Sesto Fiorentino, Firenze, Italy }
\affiliation {Physik-Institut, University of Zurich, Winterthurerstrasse 190, 8057 Zurich, Switzerland }
\affiliation {American University, Washington, D.C. 20016, USA }
\affiliation {University of Bia{\l }ystok, 15-424 Bia{\l }ystok, Poland }
\affiliation {University of Southampton, Southampton SO17 1BJ, United Kingdom }
\affiliation {University of Washington Bothell, 18115 Campus Way NE, Bothell, WA 98011, USA }
\affiliation {Institute of Applied Physics, Nizhny Novgorod, 603950, Russia }
\affiliation {Korea Astronomy and Space Science Institute, Daejeon 34055, Korea }
\affiliation {Inje University Gimhae, South Gyeongsang 50834, Korea }
\affiliation {National Institute for Mathematical Sciences, Daejeon 34047, Korea }
\affiliation {Institute of Mathematics, Polish Academy of Sciences, 00656 Warsaw, Poland }
\affiliation {Hillsdale College, Hillsdale, MI 49242, USA }
\affiliation {Hanyang University, Seoul 04763, Korea }
\affiliation {Seoul National University, Seoul 08826, Korea }
\affiliation {NASA Marshall Space Flight Center, Huntsville, AL 35811, USA }
\affiliation {ESPCI, CNRS, F-75005 Paris, France }
\affiliation {Southern University and A\&M College, Baton Rouge, LA 70813, USA }
\affiliation {College of William and Mary, Williamsburg, VA 23187, USA }
\affiliation {Centre Scientifique de Monaco, 8 quai Antoine Ier, MC-98000, Monaco }
\affiliation {Indian Institute of Technology Madras, Chennai 600036, India }
\affiliation {IISER-Kolkata, Mohanpur, West Bengal 741252, India }
\affiliation {Whitman College, 345 Boyer Avenue, Walla Walla, WA 99362 USA }
\affiliation {Indian Institute of Technology Bombay, Powai, Mumbai, Maharashtra 400076, India }
\affiliation {Scuola Normale Superiore, Piazza dei Cavalieri 7, I-56126 Pisa, Italy }
\affiliation {Universit\'e de Lyon, F-69361 Lyon, France }
\affiliation {Hobart and William Smith Colleges, Geneva, NY 14456, USA }
\affiliation {OzGrav, Swinburne University of Technology, Hawthorn VIC 3122, Australia }
\affiliation {Janusz Gil Institute of Astronomy, University of Zielona G\'ora, 65-265 Zielona G\'ora, Poland }
\affiliation {University of Washington, Seattle, WA 98195, USA }
\affiliation {King's College London, University of London, London WC2R 2LS, United Kingdom }
\affiliation {Indian Institute of Technology, Gandhinagar Ahmedabad Gujarat 382424, India }
\affiliation {Indian Institute of Technology Hyderabad, Sangareddy, Khandi, Telangana 502285, India }
\affiliation {International Institute of Physics, Universidade Federal do Rio Grande do Norte, Natal RN 59078-970, Brazil }
\affiliation {Andrews University, Berrien Springs, MI 49104, USA }
\affiliation {Universit\`a di Siena, I-53100 Siena, Italy }
\affiliation {Trinity University, San Antonio, TX 78212, USA }
\affiliation {Van Swinderen Institute for Particle Physics and Gravity, University of Groningen, Nijenborgh 4, 9747 AG Groningen, The Netherlands }
\affiliation {Abilene Christian University, Abilene, TX 79699, USA }
\affiliation {Colorado State University, Fort Collins, CO 80523, USA }



\begin{abstract}
  We report on a new all-sky search for periodic gravitational waves in the frequency band 475--2000\,Hz 
  and with a frequency time derivative in the range of $\sci{[-1.0, +0.1]}{-8}$\,Hz/s. Potential signals could be produced by
  a nearby spinning and slightly non-axisymmetric isolated neutron star in our galaxy.
  This search uses the data from Advanced LIGO's first observational run O1.
  No gravitational wave signals were observed, and upper limits were placed on their strengths. For completeness, results from the separately published low frequency search 20--475\,Hz are included as well.
  Our lowest upper limit on worst-case (linearly polarized) strain amplitude $h_0$ is  $\sim\sci{4}{-25}$ near 170\,Hz,
  while at the high end of our frequency range we achieve a worst-case upper limit of $\sci{1.3}{-24}$.
  For a circularly polarized source (most favorable orientation), the smallest
  upper limit obtained is  $\sim\sci{1.5}{-25}$.
\end{abstract}

%
%
\maketitle

\section{Introduction}
\label{sec:introduction}

In this paper we report the results of an all-sky, multi-pipeline search for continuous, nearly monochromatic gravitational waves in data from Advanced LIGO's first observational run (O1) \cite{aligo}. The search covered signal frequencies from 475\,Hz through 2000\,Hz and frequency derivatives over the range $\sci{[-1.0, +0.1]}{-8}$\,Hz/s.

A number of searches for periodic gravitational waves from isolated neutron stars
have been carried out previously in LIGO and Virgo data~\cite{S1Paper, S2TDPaper, S3S4TDPaper, Crab, S5TDPaper, CasA, S6CasA, S6SNRPaper, S6GlobularCluster,  S2FstatPaper, S2Hough, S4IncoherentPaper, S4EH, EarlyS5Paper, S5EH, FullS5Semicoherent, FullS5EH, S6PowerFlux, S6BucketEH, S6BucketFU, VSR24Hough, orionspur, O1EH, O1LowFreq, S5HF, S5Hough, VSR1TDFstat, O1Directed, O1DirectStochasticPaper}.
These searches have included coherent searches for continuous wave (CW) gravitational radiation from known radio and X-ray pulsars, directed searches for known stars or locations having unknown signal frequencies, and spotlight or all-sky
searches for signals from  unknown sources. None of those searches have found any signals, establishing limits on strength of any putative signals. No previous search for continuous waves covered the band 1750-2000\,Hz.

Three search methods were employed to analyze O1 data:
\begin{itemize}
 \item The {\em PowerFlux} pipeline has been used in previous searches of
   LIGO's S4, S5 and S6 and O1 runs~\cite{S4IncoherentPaper,EarlyS5Paper,FullS5Semicoherent,S6PowerFlux, O1LowFreq}
  and uses a {\em Loosely Coherent} method for following up outliers~\cite{loosely_coherent}. A new {\em Universal} statistic \cite{universal_statistics} provides correct upper limits
  regardless of the noise distribution of the underlying data, while still showing close to optimal
  performance for Gaussian data. 
  
  The followup of outliers uses a newly implemented dynamic programming algorithm similar to the Viterbi
  method~\cite{viterbimethod} implemented in another recent CW search of Scorpius X-1~\cite{viterbi}.

  \item The {\em SkyHough} pipeline has been used in previous all-sky searches 
of the initial LIGO S2, S4 and S5 and Advanced LIGO  O1 data
\cite{S2Hough,S4IncoherentPaper,S5Hough,O1LowFreq}. 
The use of the Hough algorithm makes it more robust than other methods with respect to noise spectral disturbances 
and phase modelling of the signal \cite{S4IncoherentPaper,AllSkyMDC}.
  Population-based frequentist upper limits are derived from the estimated average sensitivity depth  obtained 
by adding simulated signals into the data. 

\item The \td\ pipeline has been used in the all-sky searches of the Virgo VSR1 data~\cite{VSR1TDFstat}
and of the low frequency part of the LIGO O1 data~\cite{O1LowFreq}.
The core of the pipeline is a coherent analysis of narrow-band time-domain sequences 
with the {\Fstat} method~\cite{jks}. Because of heavy computing requirements of the coherent search,
the data are divided into time segments of a few days long
which are separately coherently analyzed with the {\Fstat}. 
This is followed by a search for coincidences among candidates found
in different short time segments (\cite{VSR1TDFstat}, Section 8), for a given band. 
In order to estimate the
sensitivity, frequentist upper limits are obtained by injecting simulated
signals into the data.

\end{itemize}

The pipelines present diverse approaches to data analysis, with coherence lengths from $1800$\,s to a few days, and different responses to line artifacts present in the data.

\begin{figure*}[htbp]
\begin{center}
  \includegraphics[width=7.2in]{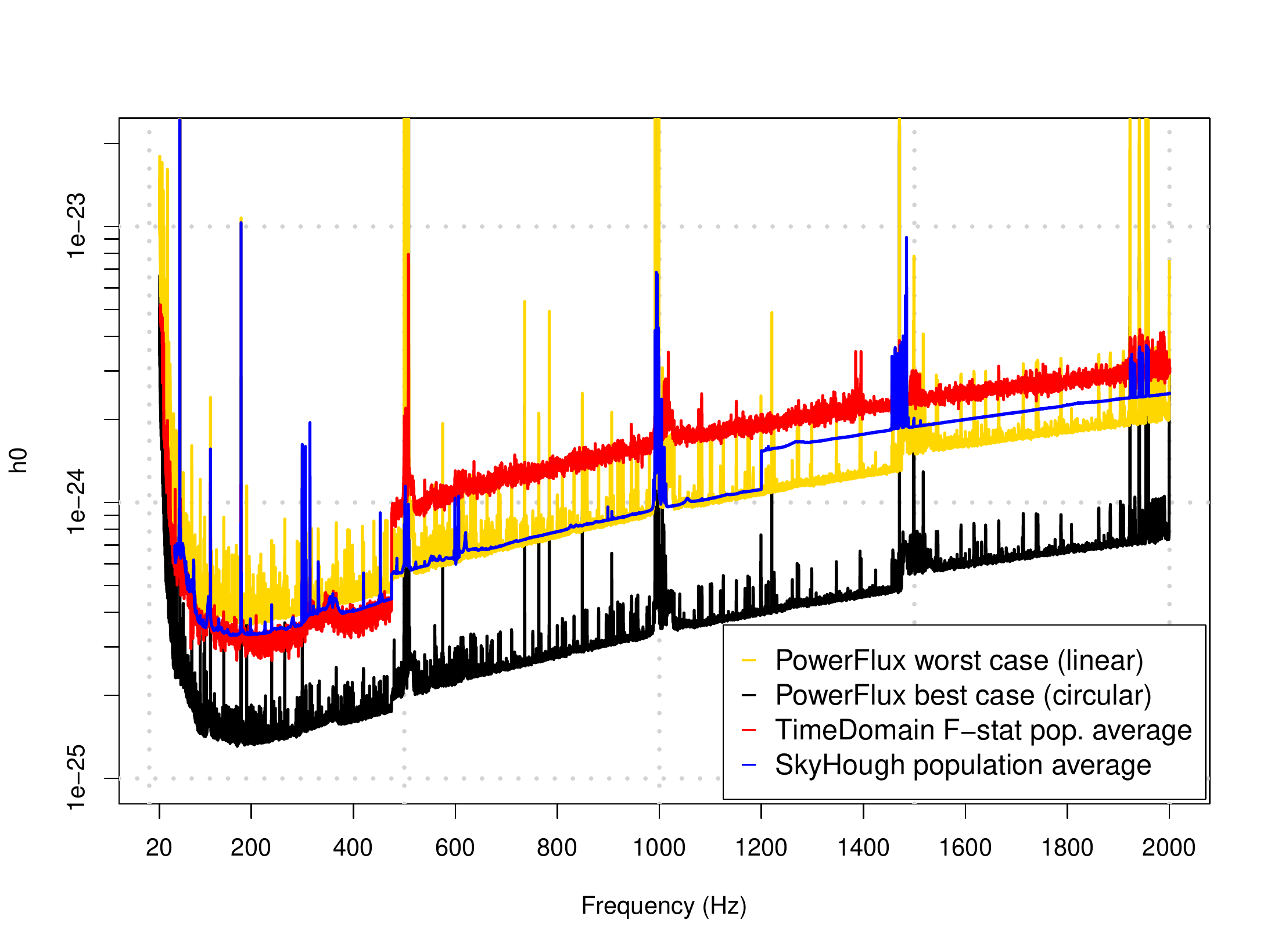}
 \caption{O1 upper limits. Looking at the right side of the plot, the upper (red) curve shows Time Domain F-statistic $95$\%~CL population averaged upper limits, the next lower curve (blue) shows maximum population average upper limits from SkyHough, folowed by yellow curve showing PowerFlux worst-case (linearly polarized) $95$\%~CL upper limits in analyzed  bands. PowerFlux upper limits are maximized over sky and all intrinsic signal parameters for each frequency band displayed. The lower (black) curve shows upper limits assuming a circularly polarized source. We include the data from the low-frequency paper \cite{O1LowFreq} to present the entire range 20--2000\,Hz. As the computational demands grow with frequencies each pipeline tuned parameters to reduce computation load. This accounts for jumps in curves at $475$, $1200$ and $1475$~Hz. The SkyHough upper limit curve shows maximum of the range of different upper limits shown in Fig. \ref{fig:skyhough_O1_upper_limits} with different upper limit values corresponding to different search depths. Because of highly non-Gaussian data the SkyHough search depths are not expected to be well-estimated for each individual search band, but are representative of the noise behaviour in the entire frequency range.
 The data for this plot can be found in \cite{data}. (color online)}
\label{fig:powerflux_O1_upper_limits}
\end{center}
\end{figure*}

After following up numerous early-stage outliers,
no evidence was found for continuous
gravitational waves in the O1 data over the band and range
of frequency derivatives searched. We therefore present bounds on detectable gravitational radiation in the form of 95\% confidence level upper limits (Fig.~\ref{fig:powerflux_O1_upper_limits}) for worst-case (linear) polarization. The worst case upper limits apply to any combination of parameters covered by the search. Best-case (circular) upper limits are presented as well, allowing one to compute the maximum distance to detected objects, under certain assumptions. Population average upper limits are produced by {\em SkyHough} and {\em \td} pipelines.

\section{LIGO interferometers and O1 observing run}
\label{sec:LIGO-O1}

The LIGO gravitational wave network consists of two observatories, one in Hanford, Washington and the other in Livingston, Louisiana, separated by a 3000-km baseline. During the O1 run each site housed one suspended interferometer with 4\,km long arms.
The interferometer mirrors act as test masses, and the passage of a gravitational wave induces a differential arm length change that is proportional to the gravitational-wave strain amplitude.
The Advanced LIGO~\cite{ALIGODescription} detectors came online in September 2015 after a major upgrade. 
While not yet operating at design sensitivity, both detectors reached an instrument noise
3 to 4 times lower than ever measured before in their most sensitive frequency band between
100 Hz and 300 Hz~\cite{DetectorPaper}.

The suspension systems of the optical elements was greatly improved, extending the usable frequency range down to 20\,Hz. Use of monolithic suspensions provided for sharper resonances of so-called {\em violin modes}, resulting in narrower (in frequency) detector artifacts. An increase in mirror
 mass has shifted the resonances to the vicinity of 500\,Hz, opening up previously-contaminated frequency bands.

With these positive effects came some new difficulties: the increase in the number of optical elements resulted in more violin modes, as well as new less well-understood resonances \cite{O1LowFreq}. 

Advanced LIGO's first observing run occurred between September 12, 2015 and
January 19, 2016, from which approximately 77 days and 66 days of analyzable
data were produced by the Hanford (H1) and Livingston (L1) interferometers, respectively.
Notable instrumental contaminants affecting the searches described here
included spectral combs of narrow lines in both interferometers, many of
which were identified after the run ended and mitigated for future runs.
These artifacts included an 8-Hz comb in H1 with the even harmonics (16-Hz comb)
being especially strong. This comb was later tracked down to digitization roundoff
error in a high-frequency excitation applied to servo-control the cavity length
of the Output Mode Cleaner (OMC). Similarly, a set of lines found to be linear
combinations of 22.7 Hz and 25.6 Hz in the L1 data was tracked down to OMC
excitation at a still higher frequency, for which digitization error occurred.

A subset of these lines with common origins at the two observatories contaminated the O1 search for a stochastic
background of gravitational waves, which relies upon cross-correlation of
H1 and L1 data, requiring excision of affected bands~\cite{O1StochasticPaper, O1DirectStochasticPaper, O1LinesPaper}.

Although most of these strong and narrow lines are stationary in frequency
and hence do not exhibit the Doppler modulations due to the Earth's motion
expected for a CW signal from most sky locations, the lines pollute the spectrum
for such sources. In sky locations near the ecliptic poles, where a putative CW signal would have little
 Doppler modulation, the lines contribute
extreme contamination for certain signal frequencies. This effect was particularly severe for the low-frequency results in the 20--475\,Hz range \cite{O1LowFreq}.

\section{Signal waveform}
\label{sec:SignalWaveform}

In this paper we assume a standard model of a spinning non-axisymmetric neutron star. Such a neutron star radiates circularly-polarized gravitational radiation along the rotation axis and linearly-polarized radiation in the directions perpendicular to the rotation axis. For the purposes of detection and establishing upper limits the linear polarization is the worst case, as such signals contribute the smallest amount of power to the detector.

The strain signal template measured by a detector is assumed to be 
\begin{equation}
\begin{array}{l}
h(t)=h_0\left(F_+(t, \alpha_0, \delta_0, \psi)\frac{1+\cos^2(\iota)}{2}\cos(\Phi(t))+\right.\\
\quad\quad\quad \left.\vphantom{\frac{1+\cos^2(\iota)}{2}}+F_\times(t, \alpha_0, \delta_0, \psi)\cos(\iota)\sin(\Phi(t))\right)\ec
\end{array}
\label{eq:h_of_t}
\end{equation}

\noindent where $F_+$ and $F_\times$ characterize the detector responses to signals with ``$+$'' and ``$\times$'' 
quadrupolar polarizations \cite{S4IncoherentPaper, EarlyS5Paper, FullS5Semicoherent}, the sky location is described by right ascension $\alpha_0$ and declination $\delta_0$, the inclination of the source rotation axis to the line of sight is denoted $\iota$, and
we use $\psi$ to denote the polarization angle (i.e. the projected source rotation axis in the sky plane).

The phase evolution of the signal is given by the formula
\begin{equation}
\label{eqn:phase_evolution}
\Phi(t)=2\pi\left(f_\textrm{source}\cdot (t-t_0)+\fdot\cdot (t-t_0)^2/2\right)+\phi\ec
\end{equation}
with $f_\textrm{source}$ being the source frequency and $\fdot$ denoting the first frequency derivative (which, when negative, is termed the {\em spindown}). 
We use $t$ to denote the time in the Solar System barycenter frame. The initial phase
$\phi$ is computed relative to reference time $t_0$.  When expressed as a function of local time of ground-based detectors, Equation \ref{eqn:phase_evolution} acquires  sky-position-dependent Doppler shift terms.

Most natural ``isolated'' sources are expected to have negative first frequency derivative, as the energy lost in gravitational or electromagnetic waves would make the source spin more slowly. The frequency derivative can be positive when the source is affected by a strong slowly-variable Doppler shift, such as due to a long-period orbit.

\section{PowerFlux search for continuous gravitational radiation}
\subsection{Overview}

This search has two main components. First, the main {\em PowerFlux} algorithm \cite{S4IncoherentPaper, EarlyS5Paper, FullS5Semicoherent, PowerFluxTechNote, PowerFlux2TechNote, PowerFluxPolarizationNote} is run to establish upper limits and produce lists of outliers with signal-to-noise ratio (SNR) greater than 5. Next, the {\em Loosely Coherent} detection pipeline \cite{loosely_coherent, loosely_coherent2, FullS5Semicoherent} is used to reject or confirm collected outliers. 

Both algorithms calculate power for a bank of signal model templates and compute upper limits and signal-to-noise ratios for each template based on comparison to templates with nearby frequencies and the same sky location and spindown. The input time series is broken into $50$\%-overlapping  long segments with durations shown in Table~\ref{tab:PowerFlux_followup_parameters}, which are then Hann-windowed and Fourier-transformed. The resulting {\em short Fourier transforms} (SFTs) are arranged into an input matrix with time and frequency dimensions. The power calculation can be expressed as a bilinear form of the input matrix $\left\{a_{t,f}\right\}$:

\begin{equation}
P[f] = \sum_{t_1, t_2} a_{t_1, f+\delta f(t_1)} a_{t_2, f+\delta f(t_2)}^* K_{t_1, t_2, f}
\end{equation}
Here $\delta f(t)$ denotes the detector frame frequency drift due to the effects from both Doppler shifts and the first frequency derivative. The sum is taken over all times $t$ corresponding to the midpoint of the short Fourier transform time interval. The kernel $K_{t_1, t_2, f}$ includes the contribution of time-dependent SFT weights, antenna response, signal polarization parameters, and relative phase terms \cite{loosely_coherent, loosely_coherent2}.

The main semi-coherent PowerFlux algorithm uses a kernel with main diagonal terms only that is easy to make computationally efficient. The {\em Loosely Coherent} algorithms increase coherence time while still allowing for controlled deviation in phase \cite{loosely_coherent}. This is done using more complicated kernels that increase effective coherence length. 

The effective coherence length is captured in a parameter $\delta$,
which describes the amount of phase drift that the kernel allows between SFTs, with $\delta=0$ corresponding to a fully coherent case, and $\delta=2\pi$ corresponding to incoherent power sums.

Depending on the terms used, the data from different interferometers can be combined incoherently (such as in stage 0, see Table~\ref{tab:PowerFlux_followup_parameters}) or coherently (as used in stages 2 or 3). The coherent combination is more computationally expensive but provides much better parameter estimation.

The upper limits (Fig.~\ref{fig:powerflux_O1_upper_limits})  are reported in terms of the worst-case value of $h_0$ (which applies to linear polarizations with $\iota=\pi/2$) and for the most sensitive circular polarization ($\iota=0$~or~$\pi$).
As described in the previous paper \cite{FullS5Semicoherent}, the pipeline does retain some sensitivity, however, to non-general-relativity GW polarization models, including a longitudinal component, and to slow amplitude evolution. A search for non-general-relativity GW signals from known pulsars is described in \cite{O1DirectedTensorial}.

The 95\% confidence level upper limits (see Fig.~\ref{fig:powerflux_O1_upper_limits}) produced in the first stage are based on the overall noise level and largest outlier in strain found for every combination of sky position, spindown, and polarization in each frequency band in
the first stage of the pipeline. These bands are analyzed by separate instances of PowerFlux \cite{FullS5Semicoherent}, and their widths vary depending on the frequency range (see Table~\ref{tab:PowerFlux_followup_parameters}).  
A followup search for detection is carried out for high-SNR outliers found in the first stage. 


\subsection{Universal statistics}

The improvements in detector noise for Advanced LIGO included extension of the usable band down to $\sim$20 Hz,
allowing searches for lower-frequency sources than previously possible with LIGO data. As discussed above,
however, a multitude of spectral combs contaminated the data, and in contrast to the 23-month
S5 Science Run and 15-month S6 Science Runs of initial LIGO, the 4-month O1 run did not span the Earth's
full orbit, which means the Doppler shift magnitudes from the Earth's motion are reduced, on the whole,
compared to those of the earlier runs. In particular, for certain combinations of sky location, frequency, and
 spindown, a signal can appear relatively stationary in frequency in
 the detector frame of reference, with the effect being most pronounced
 for low signal frequencies as noted in \cite{O1LowFreq}.

\begin{figure}[htbp]
\begin{center}
  \includegraphics[width=3.0in]{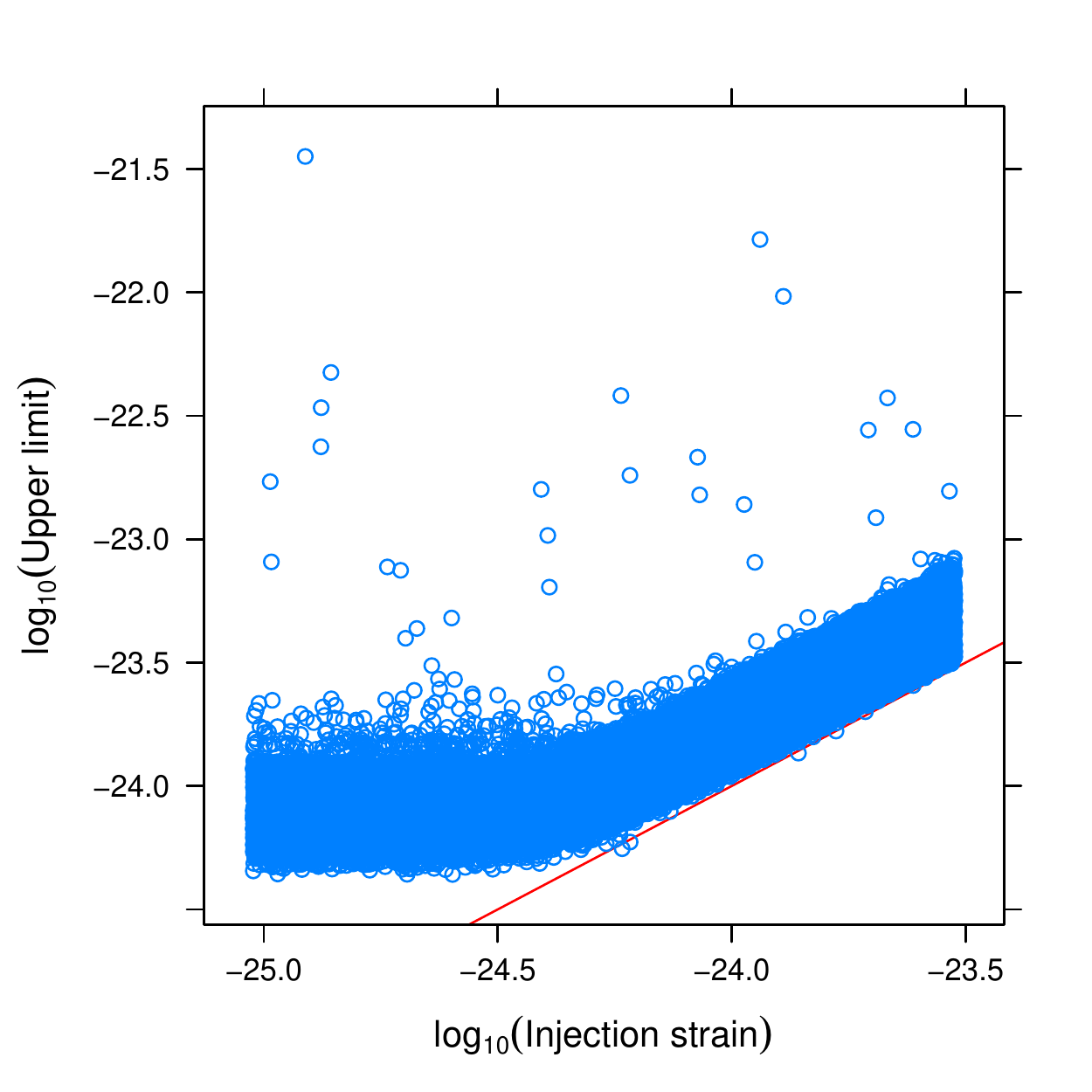}
 \caption[Upper limit versus injected strain]{PowerFlux upper limit validation. Each point represents a separate injection in the 475-1475\,Hz frequency range. Each established upper limit (vertical axis) is compared against the injected strain value (horizontal axis, red line).  The plot for high frequency range 1475-2000\,Hz is very similar and not included in this paper. (color online).}
\label{fig:ul_vs_strain}
\end{center}
\end{figure}

To allow robust analysis of the entire spectrum,
we use in this analysis the {\em Universal} statistic algorithm \cite{universal_statistics} for establishing upper limits. The algorithm is derived from the Markov inequality and shares its independence from the underlying noise distribution. It produces upper limits less than $5$\% above optimal in the case of Gaussian noise. In non-Gaussian bands it can report values larger than what would be obtained if the distribution were known, but the upper limits are always at least 95\% valid. Fig.~\ref{fig:ul_vs_strain} shows results of an injection run performed as described in \cite{FullS5Semicoherent}. Correctly-established upper limits lie above the red line.

\begin{table*}[htbp]
\begin{center}
\begin{tabular}{llccccc}\hline
Stage & Instrument sum & {Phase coherence} & \multicolumn{1}{c}{Spindown step} & \multicolumn{1}{c}{Sky refinement} & \multicolumn{1}{c}{Frequency refinement} & \multicolumn{1}{c}{SNR increase} \\
 & & \multicolumn{1}{c}{rad} & \multicolumn{1}{c}{Hz/s} &  &  & \multicolumn{1}{c}{\%}\\
\hline \hline
\multicolumn{7}{c}{20-475\,Hz frequency range, 7200\,s SFTs, 0.0625\,Hz frequency bands} \\
  0  & Initial/upper limit semi-coherent & NA  & $\sci{1}{-10}$ & $1$ & $1/2$ & NA \\
  1 & incoherent & $\pi/2$ & $\sci{1.0}{-10}$ & $1/4$ & $1/8$ & 20 \\
  2 & coherent & $\pi/2$ & $\sci{5.0}{-11}$ & $1/4$ & $1/8$ & 10  \\
  3 & coherent & $\pi/4$ & $\sci{2.5}{-11}$ & $1/8$ & $1/16$ & 10 \\
  4 & coherent & $\pi/8$ & $\sci{5.0}{-12}$ & $1/16$ & $1/32$ & 7 \\
 \hline
\multicolumn{7}{c}{475-1475\,Hz frequency range, 3600\,s SFTs, 0.125\,Hz frequency bands} \\
  0  & Initial/upper limit semi-coherent & NA  & $\sci{1}{-10}$ & $1$ & $1/2$ & NA \\
  1 & coherent & $\pi/2$ & $\sci{3.0}{-10}$ & $1/4$ & $1/8$ & 40 \\
  2 & coherent & $\pi/4$ & $\sci{1.5}{-10}$ & $1/8$ & $1/8$ & 12  \\
  3 & coherent & $\pi/8$ & $\sci{7.5}{-11}$ & $1/8$ & $1/16$ & 0 \\
 \hline
\multicolumn{7}{c}{1475-2000\,Hz frequency range, 1800\,s SFTs, 0.25\,Hz frequency bands} \\
  0  & Initial/upper limit semi-coherent & NA  & $\sci{1}{-10}$ & $1$ & $1/2$ & NA \\
  1 & coherent & $\pi/2$ & $\sci{3.0}{-10}$ & $1/4$ & $1/8$ & 40 \\
  2 & coherent & $\pi/4$ & $\sci{1.5}{-10}$ & $1/8$ & $1/8$ & 12  \\
  3 & coherent & $\pi/8$ & $\sci{7.5}{-11}$ & $1/8$ & $1/16$ & 8 \\
 \hline
\end{tabular}
\caption[Analysis pipeline parameters]{PowerFlux analysis pipeline parameters. Starting with stage 1, all stages used the {\em Loosely Coherent} algorithm for demodulation. The sky and frequency refinement parameters are relative to values used in the semicoherent PowerFlux search. The 7200\,s SFTs used for analysis of 20-475\,Hz range were too computationally expensive for higher frequencies and smaller 3600\,s and 1800\,s SFTs were used instead. The breakpoints 475\,Hz and 1475\,Hz breakpoints were chosen so that more computationally expensive range ends just before heavy instrumental artifacts due to violin modes of mirrors and beamsplitter.
}
\label{tab:PowerFlux_followup_parameters}
\end{center}
\end{table*}

\subsection{Detection pipeline}

The outlier follow-up used in \cite{FullS5Semicoherent, S6PowerFlux} has been extended with additional stages (see Table~\ref{tab:PowerFlux_followup_parameters}) to winnow the larger number of initial outliers, expected because of non-Gaussian artifacts and larger initial search space. This paper uses fewer stages than \cite{O1LowFreq} because of the use of a dynamic programming algorithm which allowed to proceed straight to coherent combinations of interferometer data.

The initial stage (marked 0) scans the entire sky with a semi-coherent algorithm that computes weighted sums of powers of Hann-windowed SFTs. These power sums are then analyzed to identify high-SNR outliers. A separate algorithm uses {\em Universal} statistics \cite{universal_statistics} to establish upper limits. 

The entire dataset is partitioned into three stretches of approximately equal length, and power sums are produced independently for any contiguous combinations of these stretches. As in \cite{orionspur, S6PowerFlux} the outlier identification is performed independently in each contiguous combination.

High-SNR outliers are subject to a coincidence test.  For each
outlier with $\SNR>7$ in the combined H1 and L1 data, we require there
to be outliers in the individual detector data of the same sky area that had $\SNR>5$,
matching the parameters of the combined-detector outlier within  $167$\,$\mu$Hz in frequency ($333$\,$\mu$Hz for the 1475--2000\,Hz band), and
$\sci{6}{-10}$\,Hz/s in spindown.  The combined-detector SNR is required to be above
both single-detector SNRs.

The identified outliers using combined data are then passed to a followup stage using the {\em Loosely Coherent} algorithm \cite{loosely_coherent} with progressively tighter phase coherence parameters $\delta$, and improved determination of frequency, spindown and sky location.

A new feature of this analysis is the use of a dynamic programming algorithm similar to the Viterbi method~\cite{viterbimethod,viterbi} in followup stages. The three stretches are each partitioned into four parts (forming 12 parts total). Given a sequence of parts the weighted sum is computed by combining pre-computed sums for each part, but the frequency is allowed to jump  by at most one sub-frequency bin. To save space, the weighted sums are maximized among all sequence combinations that have the same ending frequency bin. The use of dynamic programming made the computation efficient.

Because the resulting power sum is a maximum of many power sums, the statistics are slightly altered and are not expected to be Gaussian. They are sufficiently close to Gaussian, however, and the {\em Universal} statistic algorithm works well with this data, even though it was optimized for a Gaussian case. The followup stages use SNR produced by the same algorithm.

\begin{figure}[htbp]
\begin{center}
 \includegraphics[width=3.0in]{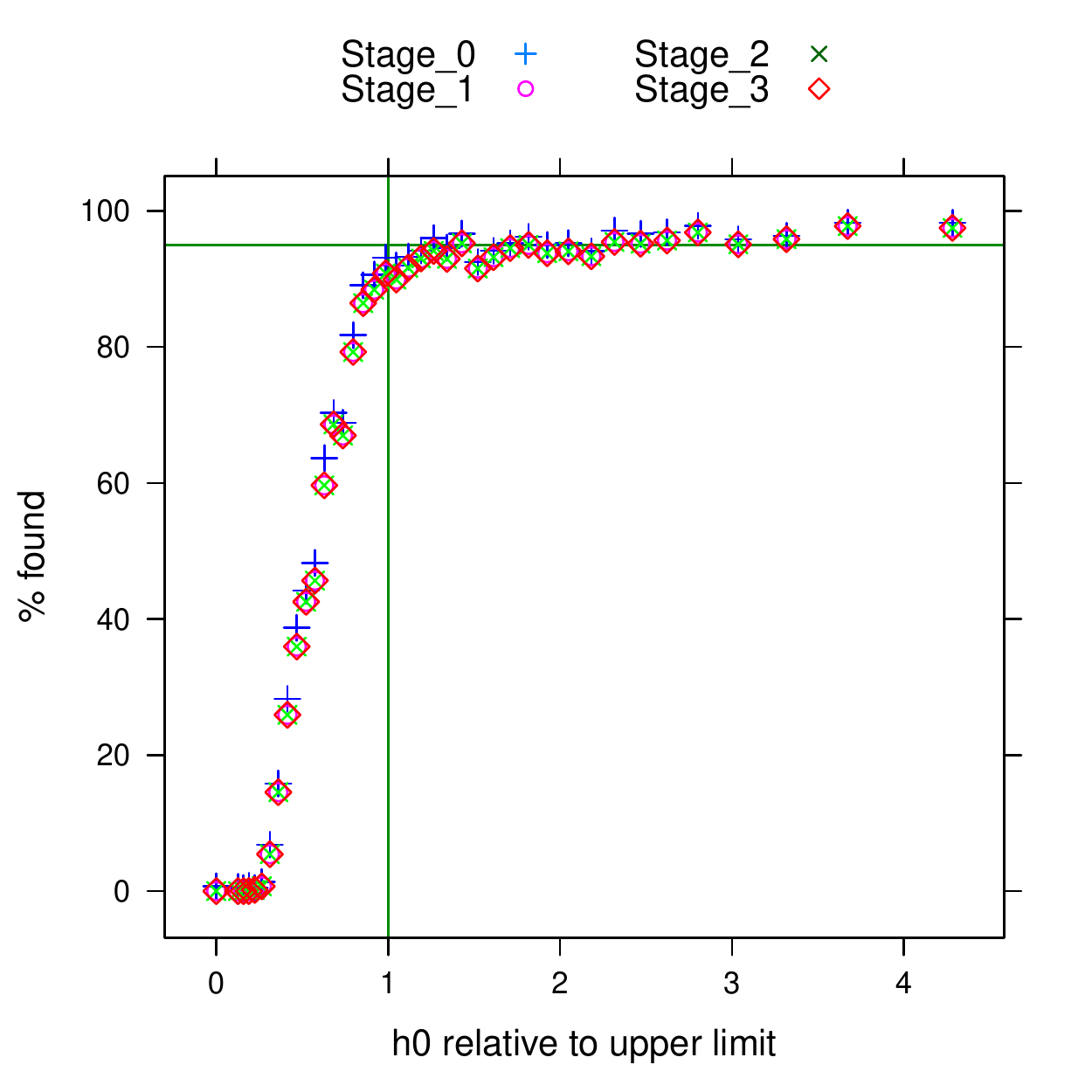}
 \caption[Injection recovery]{PowerFlux injection recovery. The injections were performed in the 475-1475\,Hz band. The injected strain divided by the upper limit in this band computed without injection is shown on the horizontal axis. The percentage of surviving injections is shown on the vertical axis, with a horizontal line drawn at the $95$\% level. Stage 0 is the output of the coincidence test after the initial semi-coherent search. The plot for high frequency range 1475-2000\,Hz is very similar and not included here. (color online). }
\label{fig:powerflux_injection_recovery}
\end{center}
\end{figure}

Allowing variation between the stretches widens the range of acceptable signals, making the search more robust. The greatest gains from this improvement, though, are in computational speed, as we can use coarser spindown steps and other parameters with only a small loss in sensitivity. This was critical to completing the Monte-Carlo simulations that verify effectiveness of the pipeline (Fig. \ref{fig:powerflux_injection_recovery}).

As the initial stage 0 sums only powers, it does not use the relative phase between interferometers, which results in some degeneracy between sky position, frequency and spindown. The first {\em Loosely Coherent} followup stage  combines
interferometer powers coherently and demands greater temporal
coherence (smaller $\delta$) , which should
boost SNR of viable outiers by at least 40\%. Subsequent stages provide tighter bounds on outlier location. Surviving outliers are passed to the Einstein@Home pipeline \cite{O1EH, S6BucketFU}.

The testing of the pipeline was performed by comprehensive simulations in each frequency range. Injection recovery efficiencies from simulations covering the $475$-$1475$\,Hz range are shown in Fig.~\ref{fig:powerflux_injection_recovery}. The simulations for higher frequencies $1475$-$2000$\,Hz produced a very similar plot, which is not shown here.

We want to highlight that simulations included highly contaminated regions such as violin modes and demonstrate the algorithm's robustness to extreme data.

In order to maintain low false dismissal rates, the followup pipeline used wide tolerances in associating outliers between stages.
 For example, when transitioning from the semi-coherent stage 0 to the {\em Loosely Coherent} stage 1, the effective coherence length increases by a factor of 4. The average true signal SNR should then increase by more than $40$\%. An additional $40$\% is expected from coherent combination of data between interferometers. But the threshold used in followup is only $40$\%, which accomodates unfavorable noise conditions, template mismatch, detector artifacts, and differences in detector duty cycle.

Our recovery criteria demand that an outlier close to the true injection location (within $3$\,mHz in frequency~$f$, $\sci{7}{-11}$\,Hz/s in spindown and [$6$\,rad$\cdot$Hz$/f$, $12$\,rad$\cdot$Hz$/f$] for [$475$-$1475$\,Hz, $1475$-$2000$\,Hz] in sky location) be found and successfully pass  through all stages of the detection pipeline. As each stage of the pipeline passes only outliers with an increase in SNR, signal injections result in outliers that strongly stand out above the background.

The followup code was verifed to recover $90$\% of injections at or above the upper limit level for a uniform distribution of injection frequencies. (Fig.~\ref{fig:powerflux_injection_recovery}). This fraction rises with injection strength. Compared with similar PowerFlux plots in earlier papers we do not reach $95$\% injection recovery right away. This is due to uneven sensitivity between interferometers (our concidence test demands an outlier be marginally seen in individual interferometers), as well as heavily contaminated data.

We note that this is still a 95\% upper limit: if a louder signal  
 had actually been present, we would have set a higher upper limit 95\% 
 of the time, even if we could only \emph{detect} the signal 90\% of the
 time.


\section{SkyHough search for continuous gravitational radiation}
\label{sec:SH_method}

\subsection{Overview}

The  {\em SkyHough} search method is described in detail in~\cite{SkyHough1, SkyHough2, Chi2Hough, S5Hough}, 
and was also used in the previous low-frequency O1 search \cite{O1LowFreq}. 
 The search  consists primarily of two main steps. First, the data from the two LIGO interferometers are  analyzed in separate all-sky searches for continuous gravitational wave signals, using a Hough transform algorithm that  produces sets of top-lists of the most significant events. In the second step, coincidence requirements on candidates are imposed.

In the first step, an implementation of the weighted Hough transform,  {\em SkyHough}~\cite{SkyHough2,S5Hough}, is used to map points from the digitized time-frequency plane of the data, called the  \emph{peak-gram},  into the space of the source parameters. The algorithm searches for signals whose frequency
evolution fits the pattern produced by the Doppler shift and spindown in the time-frequency plane of the data. In this case, the Hough number count, $n$, is the  sum of the ones and zeroes of the peak-gram weighted using the detector antenna pattern and the noise level. A useful detection statistic is the \textit{significance} (or critical ratio), and is given by
\beq
 s= \frac{n-\langle n\rangle}{\sigma} \, ,
 \eeq
 where $\langle n\rangle$ and $\sigma$ are the expected mean and standard deviation of the Hough number count for pure noise. 

The analysis  of the {\em SkyHough}  search  presented here has not identified any convincing  continuous gravitational wave signal. Hence, we proceed to set upper limits on the maximum intrinsic wave strain $h_0$ that is consistent with our observations for a population of signals described by an isolated triaxial rotating neutron star. As in previous searches, we set all-sky  population-based frequentist upper limits, that are given in different  frequency sub-bands.


\subsection{Detection pipeline}

As was done in the previous low-frequency Advanced-LIGO O1 search \cite{O1LowFreq}, covering frequencies up to 475 Hz,
 this search method uses calibrated detector $h(t)$ data to create 1800\,s Tukey-windowed SFTs, where each SFT is created from a segment of detector data that is at least 1800\,s long. From this step, 3684 and 3007 SFTs are created for H1 and L1, respectively.  SFT data from a single interferometer are analyzed by setting a threshold of 1.6 on the normalized power and then creating a peak-gram (a collection of zeros and ones).
The averaged spectrum is determined via a running-median estimation~\cite{S4IncoherentPaper} which uses 50 frequency bins to each side of the current bin.
 
The {\em SkyHough} search analyzes 0.1~Hz bands over the frequency interval 
475--2000 Hz, frequency time derivatives in the range $\sci{[-1.0, +0.1]}{-8}$~Hz/s, and covers the entire sky.
A uniform grid spacing, equal to the size of a SFT frequency bin,
$\delta f = 1/ \Tcoh= \sci{5.556}{-4}~\mathrm{Hz}$
is chosen, where $\Tcoh$ is the duration of a SFT.
The resolution in the first frequency derivative, $\delta \dot{f}$,
is given by the smallest value of $\dot{f}$ for which
the intrinsic signal frequency does not drift by more than one frequency bin during
the total observation time $\Tobs$:
$\delta \dot{f} = \delta f / \Tobs \sim \sci{4.95}{-11}~\mathrm{Hz}~\mathrm{s}^{-1}$.
This yields 203 spin-down values and 21 spin-up values for each frequency. 
%
The angular spacing of the sky grid points, $\delta \theta$ (in radians), 
is frequency dependent, with the number of templates increasing with frequency,
as given by equation (4.14) of Ref.~\cite{SkyHough1}:
\beq
\delta \theta= \frac{10^4  \, \delta f}{f N_p} \, ,
\eeq
where $N_p$ is a variable that we call pixelfactor. This variable can be manually changed to accommodate the desired sky resolution and consequently the computational cost of the search. 
The scaling factor of $10^4$ accounts for the maximum sky-position-dependent frequency modulation $v/c\sim10^{-4}$ due to
Earth's orbit.
For the Initial-LIGO S5 search $N_p$ was set to 0.5 \cite{S5Hough}, while in the previous low-frequency Advanced-LIGO O1 search \cite{O1LowFreq}
$N_p$ was set to 2, thus increasing the sky resolution by a factor of 16.

For each 0.1~Hz frequency band, the parameter space is split further into 
209 sub-regions of the sky.
For every sky region and frequency band the analysis program compiles a list of the 1000 most
significant candidates (those with the highest critical ratio values).
A final list of the 1000 most significant candidates for each 0.1~Hz frequency band is constructed, with no more than 300 candidates from a single sky region.
This procedure reduces the influence of instrumental spectral disturbances that affect specific sky regions.

As the number of  sky positions in an all-sky search  increases with the square of the frequency, the computational cost becomes larger for the highest frequencies. In order to perform this  {\em SkyHough} all-sky search within the allocated computational 
 budget, the search presented here  is split in two different bands: from 475 to 1200 Hz, and from 1200 Hz to 2000 Hz.  The pixelfactor $N_p$  is set equal to 2 for 475--1200 Hz band and equal to 0.5 for 1200--2000 Hz, thus performing a lower  sky grid resolution search at higher frequencies. Of course, these parameter choices, duration of the SFTs,  sky resolution, and size of the toplist per frequency band, have implications on the final sensitivity of the search itself compared to what could have been achieved. 
Around 1200 Hz we estimate that the sensitivity would have been 20\% better if the 
pixelfactor $N_p$ had remained 2, as can be inferred from Fig.~\ref{fig:skyhough_O1_upper_limits}.

\subsection{The post-processing stage}

The post-processing of the top-lists for each 0.1~Hz band consists of the following steps:

(i) Search for coincident candidates among the H1 and L1 data sets, using a coincidence window of $d_{\rm SH}<\sqrt{ 14}$. This dimensionless quantity is defined as:
\beq
d_{\rm SH}= \sqrt{(\Delta f  /\delta f )^2 + (\Delta \dot{f}  / \delta \dot{f}  )^2 +( \Delta \theta /\delta \theta)^2}
\eeq
to take into account the distances in frequency, spin-down and sky location with respect to the grid resolution in parameter space. 
Here $\Delta\theta$ is the sky angle separation. 
Each coincidence pair is then characterized by its harmonic mean significance value 
and a center in parameter space: 
the mean weighted value of frequency, spin-down and sky-location obtained by using their corresponding individual significance values.

(ii) The surviving coincidence pairs are clustered, using the same coincidence window of $d_{\rm SH}< \sqrt{ 14}$
applied to the coincidence centers. 
Each coincident candidate can belong to only a single cluster, and an element belongs to a cluster if there exists at least another element within that distance. 
Only the highest ranked cluster, if any,  will be selected for each 0.1~Hz band.
Clusters are ranked based on their mean significance value, but where all clusters overlapping with a known instrumental line are ranked below any cluster with no overlap. 
A cluster is always selected for each of the 0.1~Hz bands that had coincidence candidates. In most cases the cluster with the largest mean significance value coincides also with the one containing the highest individual value.

Clusters were marked if they overlapped with a list of known instrumental lines.  To perform this veto, we
 consider the frequency interval derived from 
frequency evolution given by the $f$ and $\dot{f}$ values of the center of the cluster together with its maximum  Doppler shift, and check if the resulting frequency interval overlaps with the frequency of a known line.

These steps (i)-(ii) take into account the possibility of coincidences and formation of clusters across 
boundaries of consecutive 0.1~Hz frequency bands. 

(iii) Based on previous studies~\cite{AllSkyMDC}, we require that interesting clusters must have a minimum 
population of 2; otherwise they are discarded. This is similar to the \lq\lq occupancy veto" described in \cite{BehnkePP2015}.

The remaining candidates are manually examined. In particular, outliers are also discarded if the frequency span of the cluster coincides with the list of instrumental lines described in Sec.~\ref{sec:LIGO-O1}, or if there are obvious spectral disturbances associated with one of the detectors.
Multi-detector searches, as those described in  \cite{O1LowFreq}, are also performed to verify the consistency of a possible signal, 
and surviving outliers are  passed to the Einstein@Home pipeline \cite{S6BucketFU, O1EH}.

\subsection{Upper limit computation}

As in  previous searches \cite{S5Hough, O1LowFreq}, we set a population-based frequentist upper limit at 
the 95\% confidence level.   
Upper limits are derived for each 0.1~Hz band from the  estimated average sensitivity depth, in a similar way to the
procedure used in the Einstein@Home  searches \cite{S6BucketEH, O1EH}. 

For a given signal strength $h_0$, the sensitivity depth is defined as:
\begin{equation}
 {\mathcal{D}} :={\sqrt{S_h}\over h_0}~~[ {1/\sqrt{\text{Hz}}} ].
\label{eq:sensDepth}
\end{equation} 
Here, $S_n$ is the maximum over both detectors of the 
power spectral density of the data, at the frequency of the signal.
$S_n$ is  estimated as the power-2 mean value, 
$\left( \sum_{i=1}^N  \left(S^{(i)}_{k} \right)^{-2}  /N \right)^{-2}$, across the different noise levels 
$S^{(i)}_{k}$  of the different $N$ SFTs.

Two different values of average depth are obtained for the 475--1200 Hz and 1200--2000 Hz frequency bands respectively, 
consistent with the change in the sky grid resolution during the search.
The depth values corresponding to the averaged  all-sky 95\% confidence detection efficiency are
obtained by means of simulated periodic gravitational wave signals added into the SFT data of both detectors H1 and L1 
in a limited number of frequency bands. 
In those bands, the detection efficiency, i.e., the fraction of signals that are 
considered detected, is computed as a function of signal strength $h_0$ expressed 
by the sensitivity depth.

For the 475--1200 Hz lower-frequency band,  eighteen different 0.1~Hz bands  were selected with the following starting frequencies: 
[532.4, 559.0, 580.2, 646.4, 658.5, 678.0, 740.9, 802.4, 810.2, 865.3, 872.1, 935.7, 
972.3, 976.3, 1076.3, 1081.0, 1123.4, 1186.0] Hz. 
These bands were chosen to be free of known spectral disturbances in both detectors, with no
coincidence candidates among the H1 and L1 data sets,
 and scattered over the whole frequency band.
In all these selected bands, we generated nine sets of 400 signals each, with fixed sensitivity 
depth in each set 
and random parameters $(f, \alpha,\delta,\dot f, \varphi_0, \psi, \cos \iota)$. 
Each signal was added into the data of both detectors, and  an analysis was 
done using the {\em SkyHough}  search pipeline over a frequency band of 0.1 Hz and the full spin-down range, but covering only 
one sky-patch. For this sky-patch a list of 300 loudest candidates was produced. 
Then we imposed a threshold on significance, based on the minimum significance found in the all-sky search in the 
corresponding 0.1~Hz band before any injections. 
The post-processing was then done using the same parameters used in the search, including the population veto.
A signal was considered detected if the center of the selected cluster, if any,  lay within a distance  $d_{\rm SH} < 13$
from the real injected value. This window was chosen based on previous studies \cite{AllSkyMDC} and prevented miscounts due to noise fluctuations or artifacts.

For the 1200--2000 Hz frequency band, the following eighteen different 0.1~Hz bands 
were selected: [1248.7, 1310.6, 1323.5, 1334.4, 1410.3, 1424.6, 1450.2, 1562.6, 1580.4, 1583.2, 1653.2, 
1663.6, 1683.4, 1704.3, 1738.2, 1887.4, 1953.4, 1991.5] Hz. 
The same procedure described above was applied to these bands.

\begin{figure}[htbp]
\begin{center}
\includegraphics[width=\columnwidth]{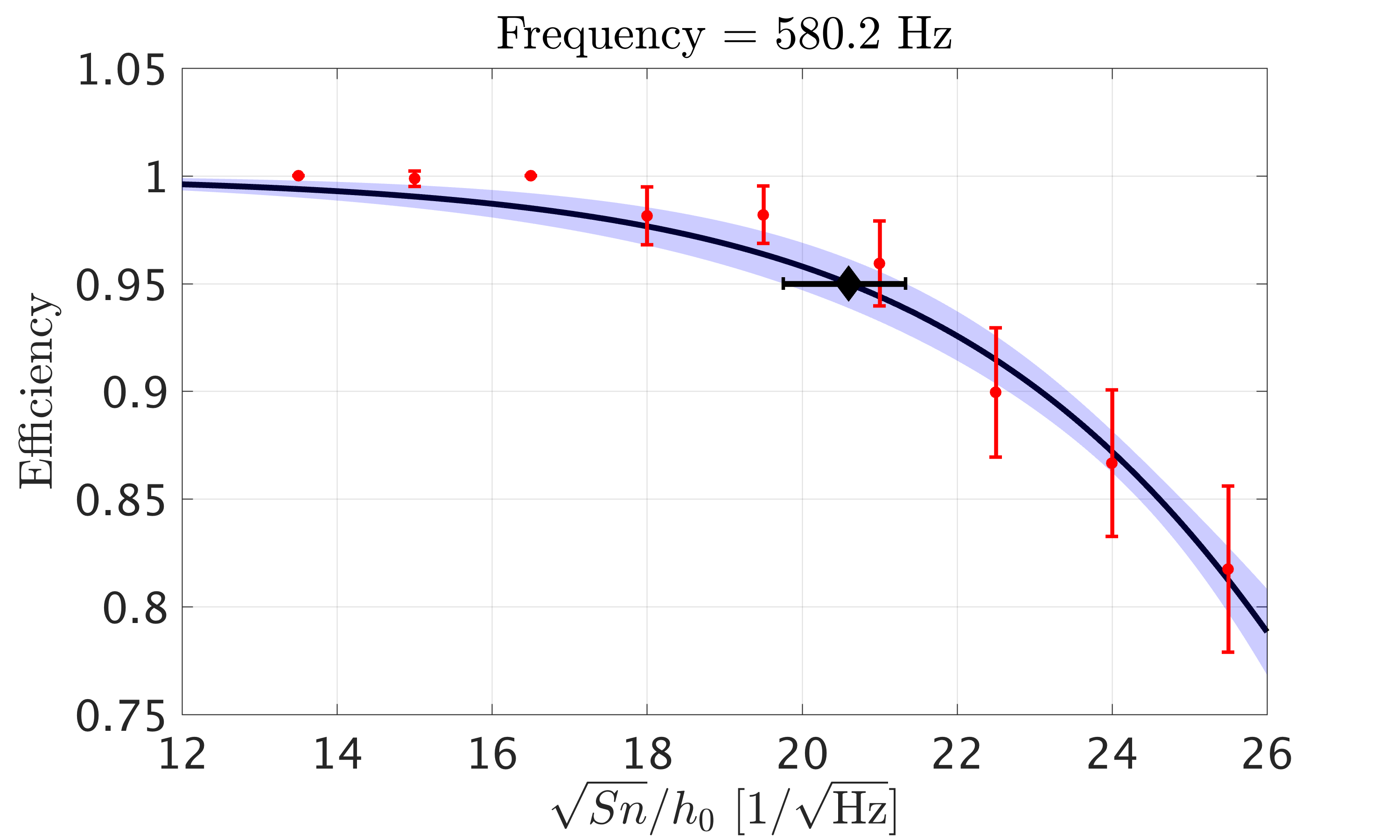}
\caption[SH_fit_16]{ Detection efficiency as a function of depth 
obtained for the 0.1 Hz frequency band starting at 580.2 Hz.
Each red dot corresponds to a set of 400 signal injections and  error bars on the data points represent the 2$\sigma_E$ standard binomial error. The (black) solid line corresponds to the fitted sigmoid curve 
and the (blue) shaded envelope  corresponds to the $2\sigma_F$ calculated according to Eq.~(\ref{eq:sigE}).
The diamond shows the depth value corresponding to the 95\% detection efficiency,  $\mathcal{D}^{95\%}$,  along with the $2\sigma_F$ uncertainty in black markers. }
\label{fig:SH_fit}
\end{center}
\end{figure}

\begin{figure}[htbp]
\begin{center}
\includegraphics[width=\columnwidth]{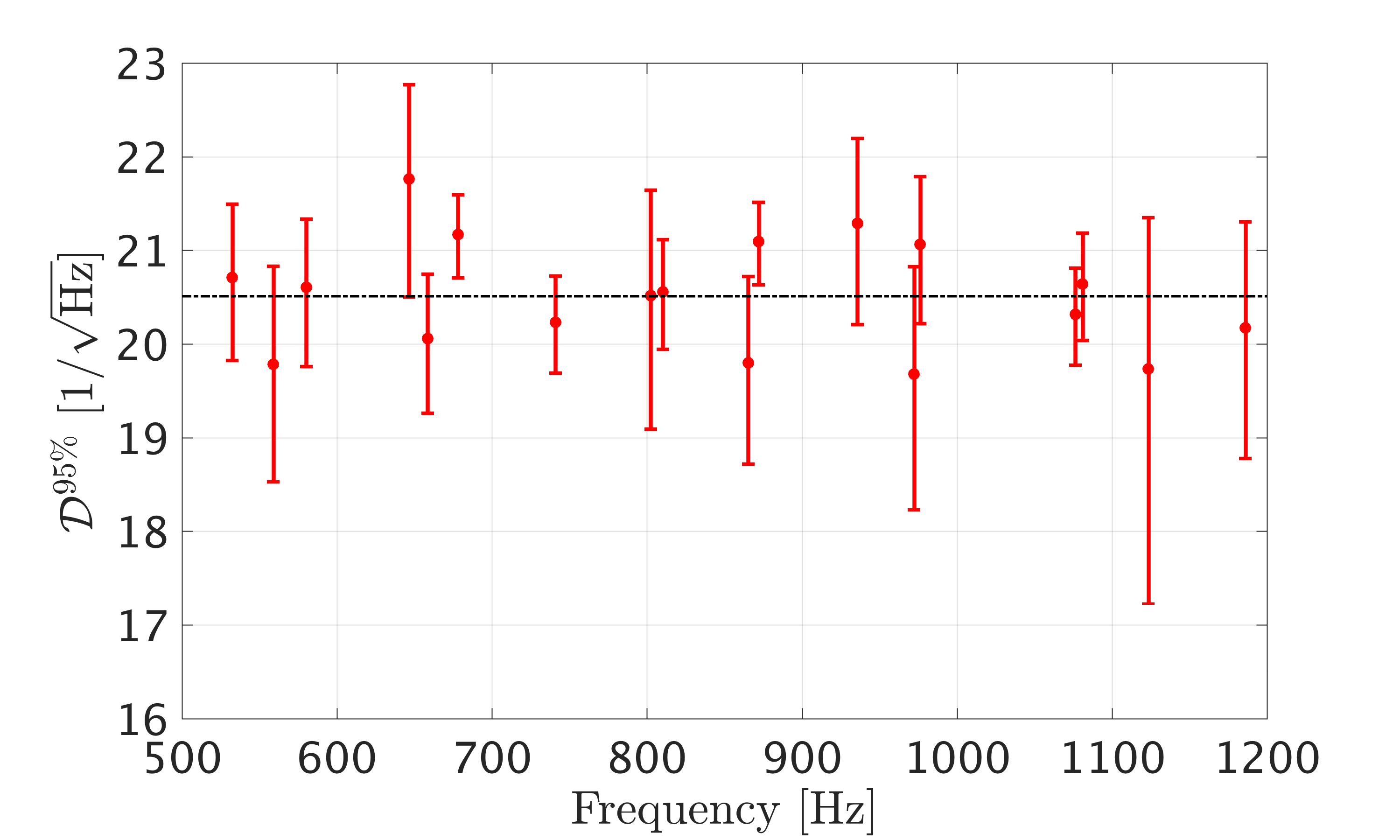}
\caption[SH_fit_low]{Depth values corresponding to the 95\% detection efficiency,  $\mathcal{D}^{95\%}$, obtained for 18 0.1 Hz frequency bands between 475 and 1200 Hz, along with their corresponding $2\sigma_F$ uncertainties  from the sigmoid fit in red markers. 
The average of the measured depths 
 at different frequencies being $\langle \mathcal{D}^{95\%} \rangle_{\rm Low} = 20.5$ $ {\text{Hz}}^{-1/2} $. 
 }
\label{fig:LowFreqDepths}
\end{center}
\end{figure}

\begin{figure}[htbp]
\begin{center}
\includegraphics[width=\columnwidth]{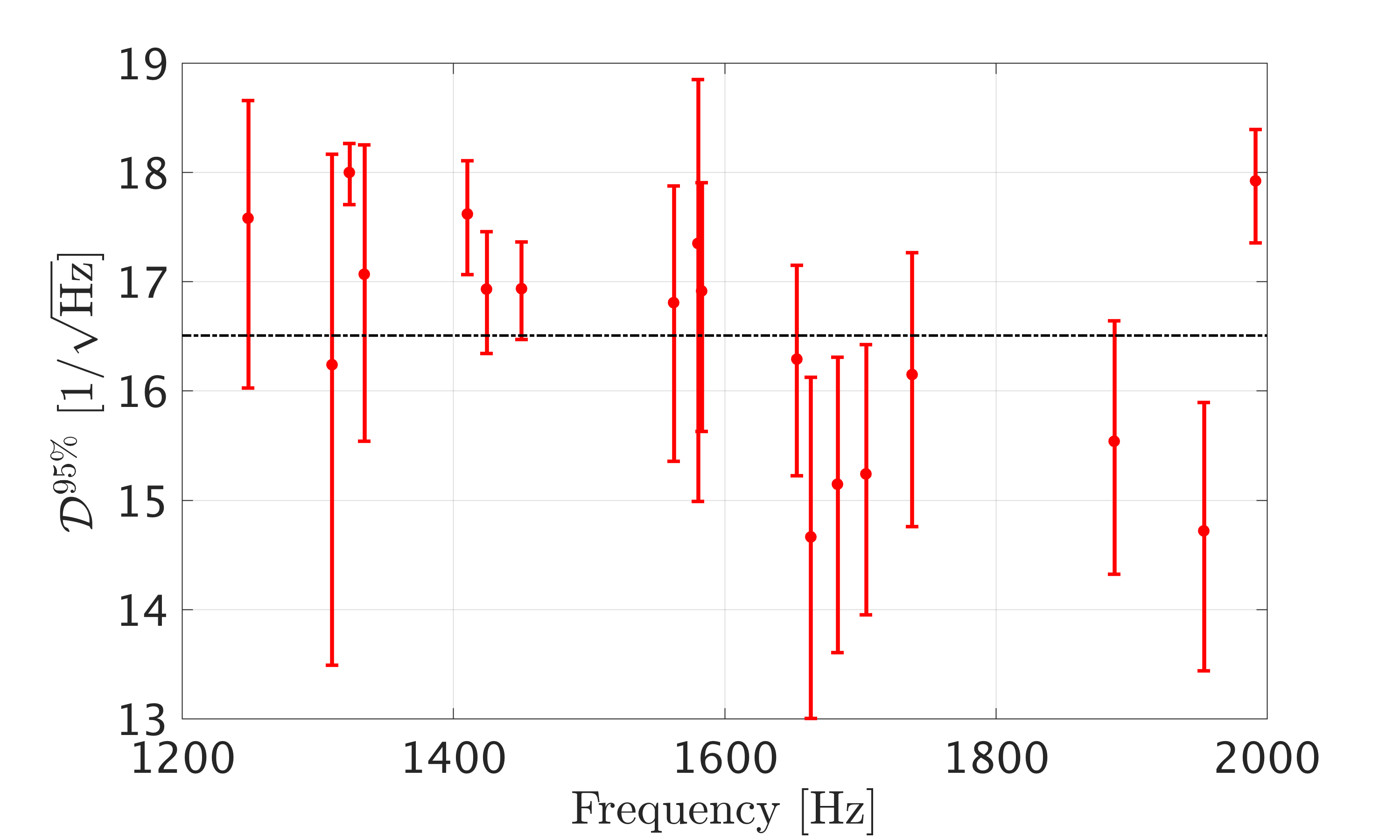}
\caption[SH_fit_high]{Depth values corresponding to the 95\% detection efficiency,  $\mathcal{D}^{95\%}$, obtained for 18 0.1 Hz frequency bands between 1200 and 2000 Hz, along with their corresponding $2\sigma_F$ uncertainties  in red markers. 
The average of the measured depths 
 at different frequencies being $\langle \mathcal{D}^{95\%} \rangle_{\rm High} = 16.5$ $ {\text{Hz}}^{-1/2} $. 
 }
\label{fig:HighFreqDepths}
\end{center}
\end{figure}

\begin{figure*}[htbp]
\begin{center}
  \includegraphics[width=7.1in]{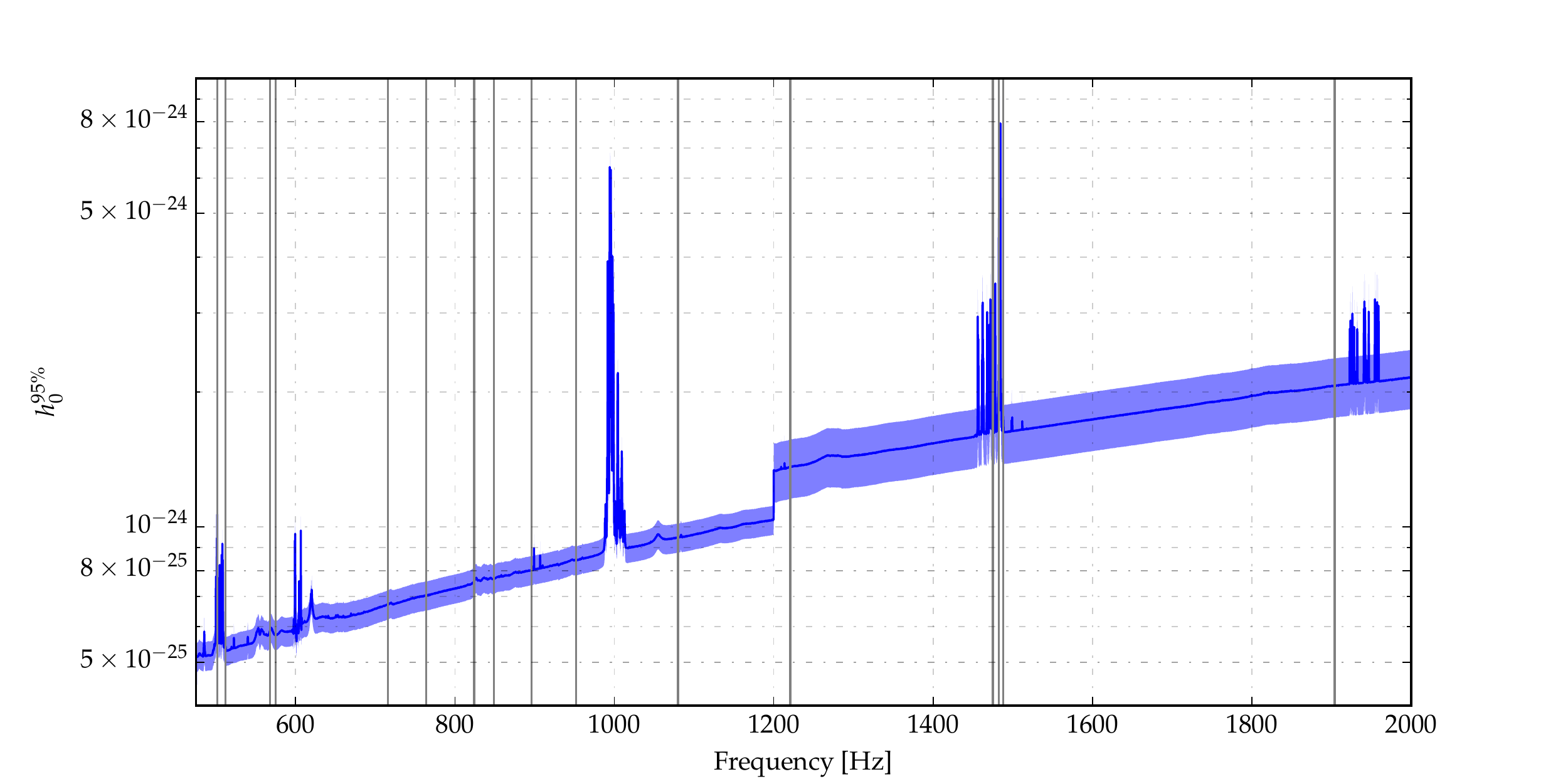}
 \caption{ {\em SkyHough} O1 upper limits. The solid (blue) line shows the averaged $95$\%~confidence level upper limits on the gravitational 
wave amplitude for every analyzed 0.1~Hz band. The vertical (grey) lines indicate 25 0.1~Hz bands in which outliers were found and consequently no upper limits were set. 
The lighter region around the upper limit represents  the $7.4\%$ and $15\%$ uncertainty levels. 
The jump in sensitivity and uncertainty at 1200 Hz corresponds to the decrease in the sky grid resolution during the search, tuned to reduce the computational load.}
\label{fig:skyhough_O1_upper_limits}
\end{center}
\end{figure*}

We collected the results from the two sets of 18 frequency bands  and for each frequency 
the  detection efficiency $E$ versus depth $\mathcal{D}$ values were fitted to a sigmoid  function  of the form:
\beq
E(\mathcal{D} ) = 1 - \frac{1}{1 + \exp(b( \mathcal{D}-a))} \, ,
\label{eq:sigmoid}
\eeq
 using the nonlinear regression algorithm  \texttt{nlinfit} provided by \texttt{Matlab}.
Since the detection rate follows a binomial distribution each data point was weighted  by the standard $\sigma_E$ error given by 
\beq
\sigma_E= \sqrt{E(1-E) \over N_I} \, ,
\eeq
where $N_I$ is the number of injections performed.
From the estimated coefficients $a$ and $b$ along with the covariance matrix $C$, we  estimated the $\sigma_F$ envelope on the fit given by
\beq
\sigma_F = \pm \sqrt{(\partial_a E)^2 C_{aa} +2 (\partial_a E )(\partial_b E) C_{ab} +(\partial_b E)^2 C_{bb}} \, ,
\label{eq:sigE}
\eeq 
where $\partial$ indicates partial derivative, and derived the corresponding depth at the 95\%  detection efficiency,  $\mathcal{D}^{95\%}$, as illustrated in Figure \ref{fig:SH_fit}.

Figures \ref{fig:LowFreqDepths} and \ref{fig:HighFreqDepths} show the obtained depth values for each  frequency corresponding to the 95\%  
efficiency level,  $\mathcal{D}^{95\%}$, together with their $2 \sigma$ uncertainty   $\delta \mathcal{D}^{95\%} = 2 \sigma_F$.

As representative of the sensitivity depth of the search, we took the average of the measured depths for each of  the  two sets of 18  different frequencies. This yielded 
$\langle \mathcal{D}^{95\%} \rangle_{\rm Low} = 20.5$ $ {\text{Hz}}^{-1/2} $
for the lower 475--1200 Hz band
and
$\langle \mathcal{D}^{95\%} \rangle_{\rm High} = 16.5$ $ {\text{Hz}}^{-1/2} $,
for the higher  1200--2000 Hz  band, 
being the range of variation observed on the measured sensitivity depth of individual frequency bands with respect to the averaged values of $7.4 \%$ and $15\%$, respectively.

The  95\% confidence upper limit on $h_0$ for undisturbed bands can then be derived by simply scaling the power spectral 
density of the data, $h_0^{95\%}= \sqrt{S_n}/\mathcal{D}^{95\%}$. The computed upper limits are shown in 
Figure~\ref{fig:skyhough_O1_upper_limits} together with their uncertainty introduced by the estimation procedure.
No limits have been placed in 25 0.1~Hz bands in which coincident candidates 
were detected, as this scaling procedure can have larger errors in those bands due to the presence of spectral disturbances.

\section{Time domain \Fstat \  search for continuous gravitational radiation}
\label{sec:TDFstat_method}

The {\td} search method uses the algorithms described in~\cite{jks,AstoneBJPK2010,VSR1TDFstat,PisarskiJ2015} and has been applied to an all-sky search of VSR1 data~\cite{VSR1TDFstat} and to the low frequency part of the LIGO O1 data~\cite{O1LowFreq}.


The main tool is the \Fstat~\cite{jks} by which one can search coherently the data over a reduced parameter space consisting of signal frequency, its derivatives, and the sky position of the source.
The F-statistic eliminates the need to sample over the four remaining parameters (see Eqs.~\ref{eq:h_of_t} and \ref{eqn:phase_evolution}): the amplitude $h_0$, the inclination angle $\iota$, the polarization angle $\psi$, and the initial phase $\phi$. Once a signal is identified the estimates of those four parameters are obtained from analytic formulae. 
However, a coherent
search over the whole 120 days long LIGO O1 data set is
computationally prohibitive and we need to apply a semi-coherent
method, which consists of dividing the data into shorter time domain
segments. The short time domain data are analyzed coherently with
the \Fstat. Then the output from the coherent search from time domain segments is analyzed by 
a different, computationally-manageable method. Moreover, to reduce 
the computer memory required to do the search, the data are divided 
into narrow-band segments that are analyzed separately. 
Thus our search method consists primarily of two parts. The first part is  the
coherent search of narrowband, time-domain segments. The second part is the
search for coincidences among the candidates obtained from the coherent search. 
The pipeline is described in Section IV 
of~\cite{O1LowFreq} (see also Figure 13 of~\cite{O1LowFreq} for the flow chart of the pipeline). 
The same pipeline is used in the high frequency analysis except that a number of parameters
of the search are different. The choice of parameters was motivated by the requirement 
to make the search computationally manageable. 

As in the low frequency search, the data are divided into overlapping frequency bands of 0.25~Hz. 
As a result, the band $[475$-$2000]$~Hz has $6300$ frequency bands. 
%
%
The time series is divided into segments, called frames, of two sidereal days long each, 
instead of six sidereal days as in the low frequency search. For O1 data, 
which is over 120 days long,
we obtain 60 time frames. Each 2-day narrowband segment contains $N = 86164$
data points. The O1 data has a number of non-science data segments. The
values of these bad data are set to zero. For this analysis, we choose only segments
that have a fraction of bad data less than 1/3 both in H1 and L1 data. 
This requirement results in twenty 2-day-long
data segments for each band. Consequently, we have $126000$ data segments to
analyze. These segments are analyzed coherently using the {\Fstat} defined by Eq.~(9) of ~\cite{VSR1TDFstat}. We
set a fixed threshold for the {\Fstat} of {$\mathcal{F}_0 = 16$} 
(in low frequency search the threshold was set to $14.5$) and record the parameters of all threshold crossings, together with the corresponding values of the signal-to-noise ratio~$\rho$,

\begin{equation}
\rho = \sqrt{2(\F - 2)}.
  \label{eq:tdfstat_O1_fstat_snr}
\end{equation}

Parameters of the threshold crossing constitute a candidate
signal.

At this first stage we also veto candidate signals overlapping with the instrumental lines identified by independent analysis of the detector data. 

For the search we use a four-dimensional grid of templates 
(parametrized by frequency, spindown, and two more parameters
related to the position of the source in the sky)
constructed in Sec.\ 4 of~\cite{PisarskiJ2015},
which belongs to the family $S_1$ of grids considered in~\cite{PisarskiJ2015}.
The grid's minimal match is $\textrm{MM}=1/2$.
It is considerably looser than in the low frequency search where the parameter 
$\textrm{MM}$ was chosen to be $\sqrt{3}/2$. 
The quality of a covering of space by lattice of identical hyperspheres
is expressed by the covering thickness $\theta$, which is defined
as the average number of hyperspheres that contain a point in the space.
In four dimensions the optimal lattice covering, i.e.\ having the minimum is called $A_4^\star$ and it has the thickness $\theta\cong 1.765529$.
The thickness of the new loose grid equals 1.767685, which 
is only $\sim$0.1\% larger than the 
 
In the second stage of the analysis we search for coincidences among the
candidates obtained in the coherent part of the analysis. We use exactly the same
coincidence search algorithm as in the analysis of VSR1 data and described in
detail in Section 8 of~\cite{VSR1TDFstat}. We search for coincidences in each of the
bands analyzed. To estimate the significance of a given coincidence, we use the
formula for the false alarm probability derived in the appendix of~\cite{VSR1TDFstat}.
Sufficiently significant coincidences are called outliers and subjected to further investigation.

The sensitivity of the search is estimated by the same procedure
as in the low frequency search paper (\cite{O1LowFreq}, Section IV). The sensitivity is taken to be the amplitude $h_0$ of the gravitational wave signal that can be confidently detected.
 We perform the following Monte-Carlo simulations.
For a given amplitude $h_0$, we randomly select the other seven
parameters of the signal: $\omega_0, \omega_1, \alpha, \delta, \phi_0, \iota$ and $\psi$.  We choose frequency and spindown parameters uniformly over their range, and source positions uniformly over the sky.  We choose angles $\phi_0$
and $\psi$ uniformly over the interval $[0, 2\pi]$ and $\cos\iota$
uniformly over the interval $[-1, 1]$. We add the signal with selected parameters to the O1 data. Then the data are processed through our pipeline.
First, we perform a coherent {\Fstat} search of each of the data segments where 
the signal was added. Then the
coincidence analysis of the candidates is performed. The signal is considered to be 
detected, if it is coincident in more than 13 of the 20 time frames analyzed for 
a given band. We repeat the simulations one hundred times.
The ratio of numbers of cases in which the signal is detected to the one hundred 
simulations performed for a given $h_0$ determines the frequentist sensitivity 
upper limits. We determine the sensitivity of the search in each of the 6300 frequency 
bands separately.
The 95\% confidence upper limits for the whole range of frequencies are given 
in Figure~\ref{fig:tdfstat_O1_upper_limits}; they follow very well the noise curves 
of the O1 data that were analyzed. The sensitivity of our high frequency search is 
markedly lower than in the low frequency search. This is because here we have a shorter 
coherent integration time, a looser grid, and a higher threshold.

\section{Search results}
\label{sec:results}

\subsection{PowerFlux results}
\label{sec:PowerFluxResults}

The PowerFlux algorithm and {\em Loosely Coherent} method compute power estimates for gravitational waves in a given frequency band for a fixed set of templates. The template parameters include frequency, first frequency derivative and sky location. The power estimates are grouped using all parameters except frequency into a set of arrays and each array is examined separately.

Since the search target is a rare monochromatic signal, it would contribute excess power to one of the frequency bins after demodulation. The upper limit on the maximum excess relative to the nearby power values can then be established. For this analysis we use a {\em Universal} statistic \cite{universal_statistics} that places conservative 95\%-confidence-level upper limits for an arbitrary statistical distribution of noise power. The implementation of the {\em Universal} statistic used in this search has been tuned to provide close-to-optimal values in the common case of Gaussian distribution.


\begin{figure}[htbp]
\includegraphics[width=3in]{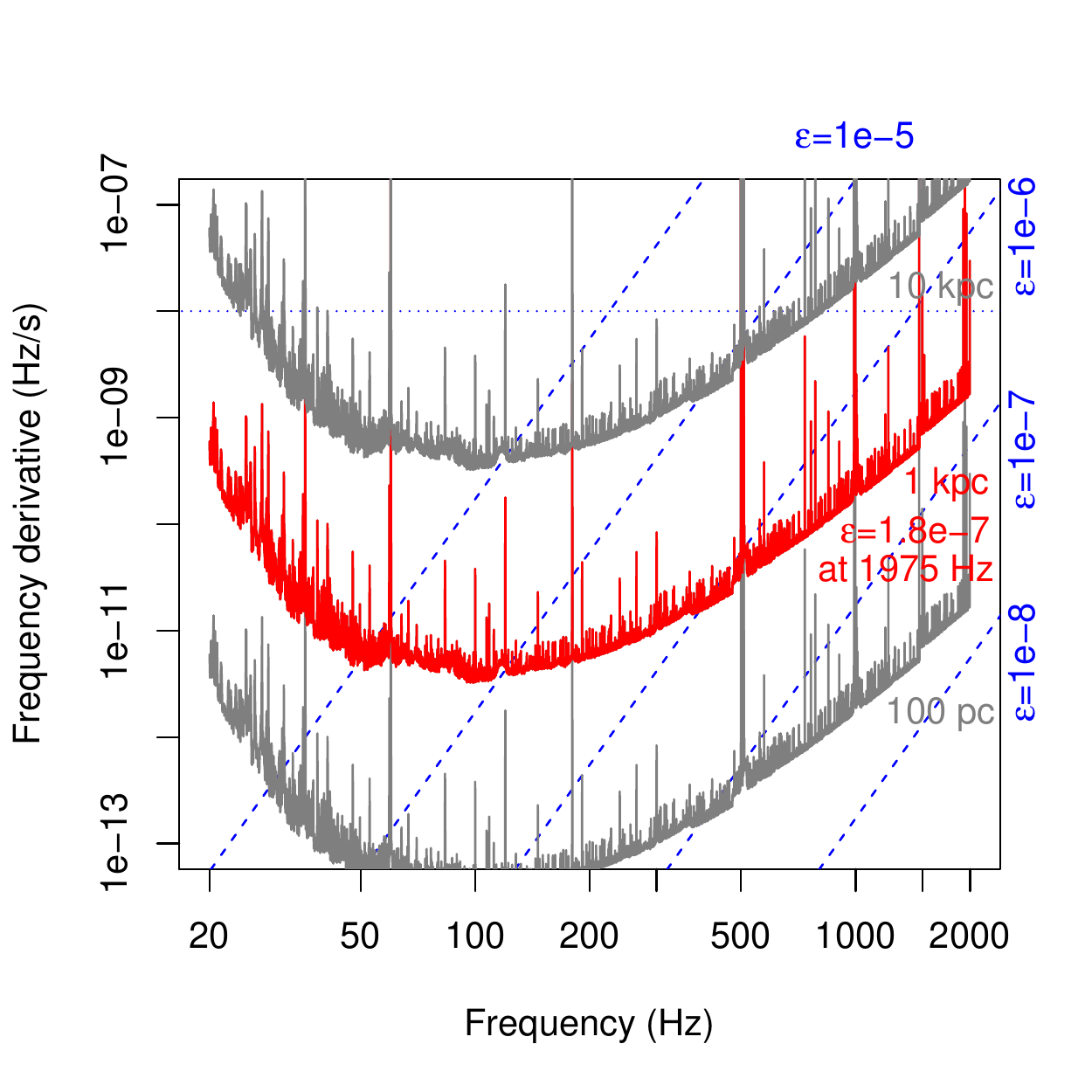}
\caption[Spindown range]{
\label{fig:spindown_range}
Range of the PowerFlux search for neutron stars
spinning down solely due to gravitational radiation.  This is a
superposition of two contour plots.  The grey and red solid lines are contours of the maximum distance at which a neutron
star could be detected as a function of gravitational-wave frequency
$f$ and its derivative $\dot{f}$.  The dashed lines 
are contours of the corresponding ellipticity
$\epsilon(f,\dot{f})$. The fine dotted line marks the maximum spindown searched. Together these quantities tell us the
maximum range of the search in terms of various populations (see text
for details) (color online).}
\end{figure}

The upper limits obtained in the search are shown in Fig.~\ref{fig:powerflux_O1_upper_limits}. The numerical data for this plot can be obtained separately \cite{data}. The upper (yellow) curve shows the upper limits for a worst-case (linear) polarization when the smallest amount of gravitational energy is projected towards Earth. The lower curve shows upper limits for an optimally oriented source. Because of the day-night variability of the interferometer sensitivity due to anthropogenic noise, the  upper limits for linearly polarized sources are more severely
affected by detector artifacts, as the detector response to linearly polarized sources varies with the same period. We are able to establish upper limits over the entire frequency range, including bands containing harmonics of 60\,Hz and violin modes.

Each point in Fig.~\ref{fig:powerflux_O1_upper_limits} represents a maximum over the sky: only small portions of the sky are excluded, near the ecliptic poles, which are highly susceptible to detector artifacts due to stationary frequency evolution produced by the combination of frequency derivative and Doppler shifts. The exclusion procedure is described in \cite{FullS5Semicoherent} and applied to $0.1$\% of the sky over the entire run.


If one assumes that the source spindown is solely due to emission of gravitational waves, then it is possible to recast upper limits on source amplitude as a limit on source ellipticity. Figure~\ref{fig:spindown_range} shows the reach of our search under different assumptions on source distance. Superimposed are lines corresponding to sources of different ellipticities.


\begin{table*}[htbp]
\begin{center}
\begin{tabular}{lD{.}{.}{6}rD{.}{.}{5}D{.}{.}{4}}
\hline
Label & \multicolumn{1}{c}{Frequency} & \multicolumn{1}{c}{Spindown} & \multicolumn{1}{c}{$\RAJ$} & \multicolumn{1}{c}{$\DECJ$} \\
 & \multicolumn{1}{c}{Hz} & \multicolumn{1}{c}{nHz/s} & \multicolumn{1}{c}{degrees} & \multicolumn{1}{c}{degrees} \\
\hline \hline
ip0   &  265.575533  & $\sci{-4.15}{-3}$   &   71.55193     &  -56.21749 \\
ip1   &  848.969641  & $\sci{-3.00}{-1}$   &   37.39385     &  -29.45246 \\
ip2   &  575.163521  & $\sci{-1.37}{-4}$   &  215.25617     &    3.44399 \\
ip3   &  108.857159  & $\sci{-1.46}{-8}$   &  178.37257     &  -33.4366  \\
ip4   & 1393.540559  & $\sci{-2.54}{-1}$   &  279.98768     &  -12.4666  \\
ip5   & 52.808324  & $\sci{-4.03}{-9}$     &  302.62664     &  -83.83914 \\
ip6   &  146.169370 & $\sci{-6.73}{0\mathrm{\;\;\;}}$     &  358.75095     &  -65.42262 \\
ip7   & 1220.555270 & $\sci{-1.12}{0\mathrm{\;\;\;}}$     &  223.42562     &  -20.45063 \\
ip8   &  191.031272 & $\sci{-8.65}{0\mathrm{\;\;\;}}$     &  351.38958     &  -33.41852 \\
ip9   &  763.847316 & $\sci{-1.45}{-8}$    &  198.88558     &   75.68959 \\
ip10  &   26.341917 & $\sci{-8.50}{-2}$    &  221.55565     &   42.87730 \\
ip11  &   31.424758 & $\sci{-5.07}{-4}$    &  285.09733     &  -58.27209 \\
ip12  &   38.477939 & $\sci{-6.25}{0\mathrm{\;\;\;}}$     &  331.85267     &  -16.97288 \\
ip13  &   12.428001 & $\sci{-1.00}{-2}$     &  14.32394     &  -14.32394 \\
ip14  & 1991.092401 & $\sci{-1.00}{-3}$     & 300.80284     &  -14.32394 \\
\hline
\end{tabular}
\caption[Parameters of hardware injections]{Parameters of the hardware-injected simulated continuous-wave signals during the O1 data run (epoch GPS $1130529362$). Because the interferometer configurations were largely frozen in a preliminary state after the first discovery of gravitational waves from a binary black hole merger, the hardware injections were not applied consistently. There were no injections in the H1 interferometer initially, and the initial injections in the L1 interferometer used an actuation method with significant inaccuracies at high frequencies.}
\label{tab:injections}
\end{center}
\end{table*}

The detection pipeline produced 31 outliers located in the 1000--1033\,Hz region heavily contaminated with violin modes (Table~\ref{tab:PowerFluxOutliersViolin}), 134 outliers spanning only one data segment (about 1 month) that are particularly susceptible to detector artifacts (Tables~\ref{tab:PowerFluxOutliersShortMid} and \ref{tab:PowerFluxOutliersShortHigh}), and  48 outliers (Table~\ref{tab:PowerFluxOutliers}) that do not fall into either of those two categories. Each outlier is identified by a numerical index. We report SNR, frequency, spindown and sky location. 

The ``Segment'' column describes the persistence of the outlier through
the data, and specifies which contiguous subset of the three equal
partitions of the timespan contributed most significantly to the
outlier: see \cite{orionspur} for details.  A true continuous signal
from an isolated source would normally have
[0,2] in this column (similar contribution from all 3 segments), or on
rare occasions [0,1] or [1,2].  Any other range is indicative of a
statistical fluctuation, an artifact or a signal that does not conform
to the phase evolution of Equation~\ref{eqn:phase_evolution}.

During the O1 run several simulated pulsar signals were injected into the data by applying a small force to the interferometer mirrors with auxiliary lasers. 
Several outliers were due to such hardware injections (Table~\ref{tab:injections}). 

The recovery of the hardware injections gives us additional confidence that no potential signal was missed. 
Manual followup has shown non-injection outliers spanning all three segments to be caused by pronounced detector artifacts. Outlier number 72 in Table~\ref{tab:PowerFluxOutliers} spanning two segments was also investigated with a fully coherent followup based on the Einstein@Home pipeline~\cite{S6BucketFU, O1EH}. No outlier was found to be consistent with the astrophysical signal model.


\subsection{SkyHough results}
\label{sec:SkyHoughResults}

In this section we report the main results of the O1 all-sky search
 between 475 and 2000 Hz using the {\em SkyHough} 
 pipeline, as described in section \ref{sec:SH_method}.
In total, 71 0.1~Hz bands contained coincidence candidates: 
19 in the 475--1200 Hz band, analysed with higher sky resolution, and 52 in the 
1200--2000 Hz band, analysed with lower sky resolution. 

After discarding all the clusters containing only one coincidence pair, this list was reduced to 25 outliers, 
17 in the low frequency band and 8 in the high frequency band, which  were further inspected.
A detailed list of these remaining outliers is shown in Table~\ref{tab:sh_outliers}.
Among the 25 outliers, 17 were related to known line artifacts contaminating either H1 or L1 data and 7 were identified with the 
 hardware injected pulsars ip1, ip2, ip7 and ip9.

\begin{table*}[htbp]
\begin{center}
\begin{tabular}{cccccc}\hline
Label & $s_{mean}$  & Frequency  & Spin-down & $\alpha$ & $\delta$  \\
 &  & [Hz] &  [nHz/s]&  [deg] &  [deg]    \\
\hline \hline \\
ip2 & 30.50 & 575.1635 (0.0001) & 0.0170 (0.0171)    & 215.1005 (0.1557) & 3.0138 (0.4302)  \\
ip9 & 35.85 & 763.8507 (0.0034) & $-$0.5567 (0.5567) & 203.8965 (5.0109) & 73.8445 (1.8451)  \\
ip1 & 36.06 & 848.9657 (0.0053) & 0.5497 (0.2497)    & 37.7549 (0.3611) & $-$25.2883 (4.1642) \\
ip7 & 41.61 & 1220.5554 (0.0009)& 0.5482 (0.5718)    & 229.2338 (5.8082) & 4.1538 (24.6044)  \\
\hline
\end{tabular}
\caption[]{{\em SkyHough} hardware injection cluster information. The table provides the frequency, 
spin-down and sky location of the cluster center related to each of the hardware injections found by the {\em SkyHough} search. In parentheses the distance from the cluster center  to the injected values are shown.
Frequencies are converted to epoch GPS 1125972653.}
\label{tab:sh_hi}
\end{center}
\end{table*}

Table~\ref{tab:sh_hi} presents the parameters of the center of the clusters obtained related to these hardware injections. 
Two hardware injection were not recovered. Ip4 was not found since its spin-down was outside the search range, and ip14 was linearly polarized 
 and had a strain amplitude $h_0$ below our sensitivity.

The only unexplained outlier around  715.7250 Hz, corresponding to Idx=6 in Table~\ref{tab:sh_outliers}, was further investigated.  
A multi-detector Hough search was performed to verify the consistency of a possible signal. 
In this case the maximum combined significance obtained was  $5.98$ while we would have expected a minimum value of $8.21$  
in case of a real signal.
The outlier was also followed up with the  Einstein@Home pipeline \cite{S6BucketFU} using coherent integration times of 210  and 500 hours. 
This search covered signal frequencies in the range $[715.724,  715.726]$\,Hz (epoch GPS 1125972653), 
frequency derivatives  over $\sci{[-2.2, -1.9]}{-9}$\,Hz/s, 
and a sky region RA = $1.063 \pm 0.020$\,rad, 
DEC = $-0.205 \pm 0.020$\,rad that included the whole associated cluster. This search showed that this candidate was not interesting and had a very low probability of having astrophysical origin. 

%

Therefore, this {\em SkyHough}  search did not find any evidence of a continuous gravitational wave signal. Upper limits
have been computed in each 0.1~Hz band, except for the 25 bands in which outliers
 were found.

%
%
%
%


\begin{table*}[htbp]
\begin{center}
\begin{tabular}{cccccc}\hline
Label & FA & Frequency [Hz] & Spin-down [nHz/s] & $\alpha$ [deg] & $\delta$ [deg] \\
\hline \hline \\
ip1 & 0 & 848.9687 (0.0007) & -2.4474 (2.1474) &  39.4542 (2.0603) & $-$39.4354 (9.9830) \\
ip2 & 0 & 575.1638 (0.0003) & 0.0162 (0.0163) &  203.8658 (11.3903) & $-$27.1485 (30.5924) \\
ip4 & 0 & 1393.5286 (0.0021) & $-$24.901 (0.5991) & 281.4735 (1.4858) & $-$13.3001 (0.8340) \footnote{Spin down of ip4 was outside the search range. The estimate was obtained by extending the spin down range in the band where the hardware injection is located.}\\
ip7 & 0 & 1220.5540 (0.0007) & $-$0.0784 ($-$1.0416) & 218.8902 (4.5354) & $-$32.1127 (11.6621) \\
ip9 & 0 & 763.8472 (0.0001) & $-$0.0503 (0.0503) & 197.8817 (1.0039) & 75.9108 (0.2212)  \\
\hline
\end{tabular}
\caption[]{Hardware injection recovery with the {\td} pipeline. The values in parentheses are the absolute errors, that is, the difference with respect to the injection parameters. Frequencies are converted to epoch GPS 1131082120.}
\label{tab:tdfstat_hi}
\end{center}
\end{table*}

\subsection{Time domain \Fstat \ results}
\label{sec:TDFstatResults}

In the $[475, 2000]$ Hz bandwidth range under study, 6300
0.25-Hz wide bands were analyzed.    
Vetoing candidates around the
known interference lines, a certain fraction of the bandwidth was not analyzed. 
As a result 26\% of the $[475, 2000]$ Hz band was vetoed, overall.

\begin{figure*}[htbp]
  \begin{center}
  \includegraphics[width=7.2in]{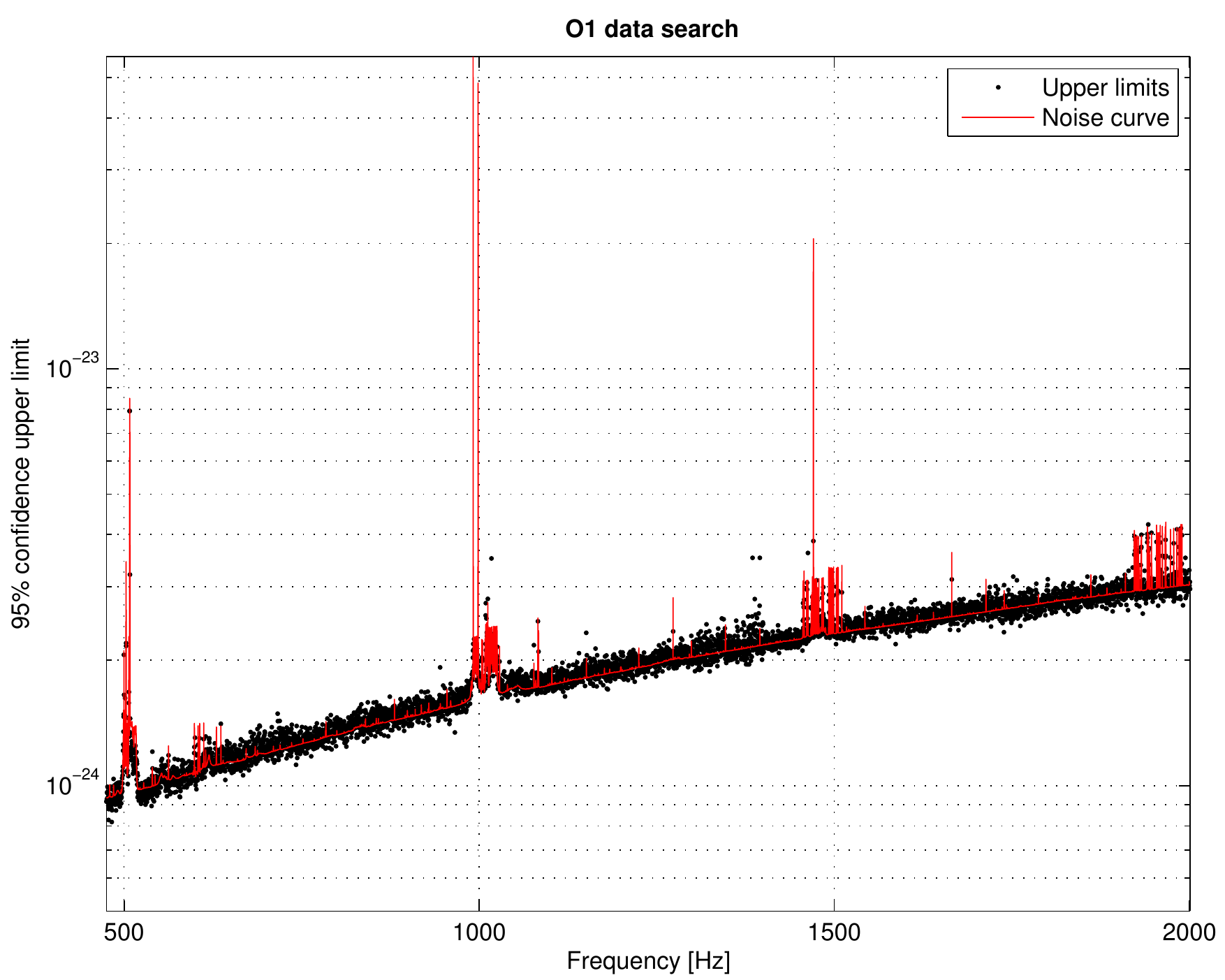}
  \caption{{\td} pipeline O1 upper limits. Black dots are the 95\% confidence upper limits for each frequency,
the red line denotes the H1 and L1 detectors' average noise curve rescaled by the factor $38/\sqrt{T_0}$,
where $T_0=172328$~s is the observational time of the 2-sidereal-day time series segment (color online). The factor of $38$ is larger than the factor of $27.5$ obtained the low frequency search indicating loss of sensitivity due to a looser grid of templates used here.}
  \label{fig:tdfstat_O1_upper_limits}
  \end{center}
\end{figure*}

\begin{figure}[htbp]
  \includegraphics[width=\columnwidth]{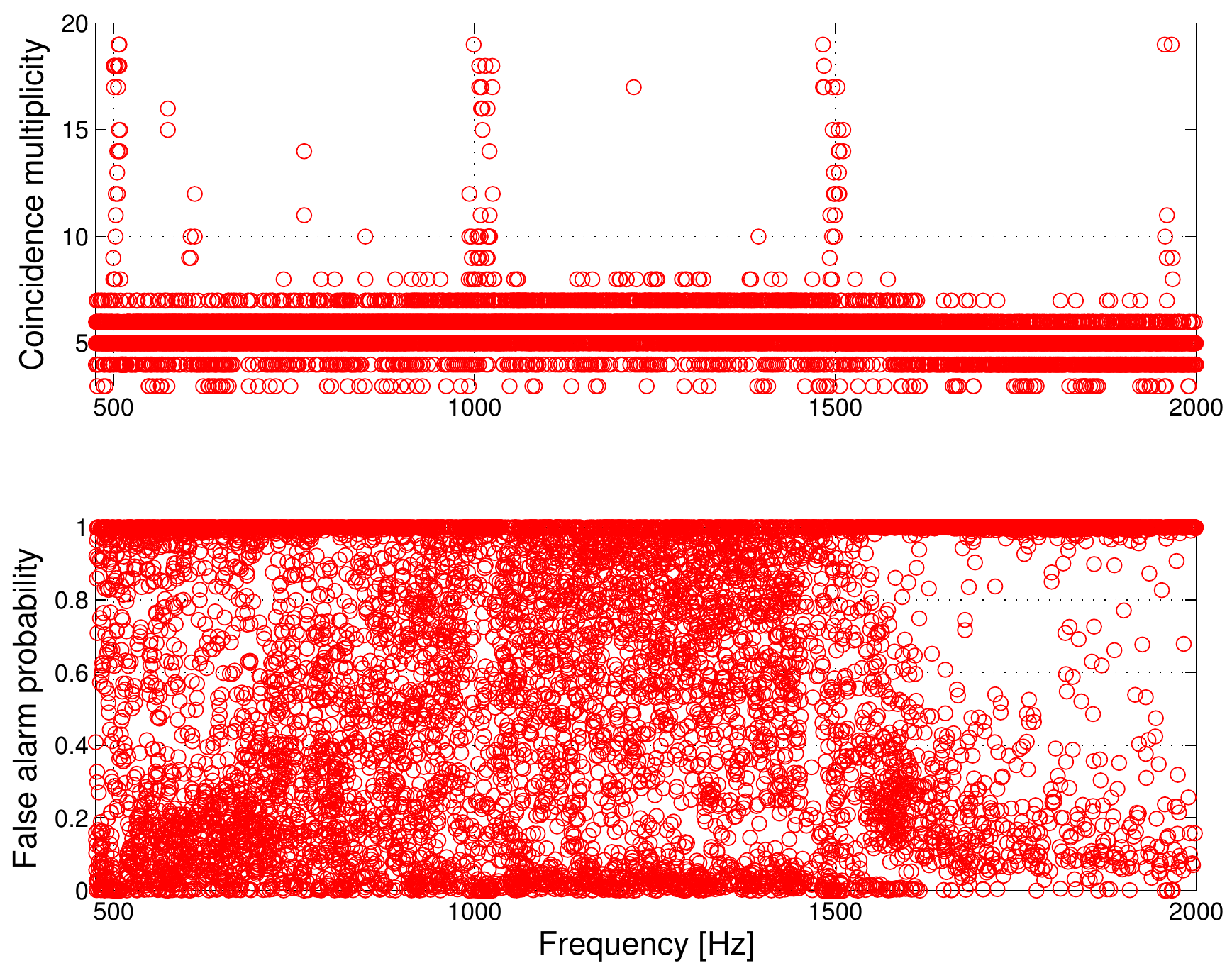}
  \caption{Results of {\td} pipeline coincidences as a function of the
band frequency. Top panel: maximum coincidence multiplicity. Bottom panel:
false alarm probability for the coincidence with the maximum multiplicity.}
  \label{fig:tdfstat_O1_Coinc_FA}
\end{figure}

Of 6300 bands analyzed, 307 bands were completely vetoed because of the line artifacts.  
As a result, the search was performed in the remaining 5993 bands. As twenty 2-days segments have been chosen for the analysis, the 119860 data segments were analyzed coherently with the \Fstat. From the coherent search we obtained around $8.6 \times 10^{10}$ candidates. These candidates were subject to a search for initial
coincidences in the second stage of the {\td} analysis. The search
for coincidences was performed in all the bands except for the above-mentioned
307 that were completely vetoed.  In the coincidence analysis, for each band, the coincidences
among the candidates were searched in twenty 2-day long time frames.  In
Figure~\ref{fig:tdfstat_O1_Coinc_FA} the results of the coincidence search are
presented. The top panel shows the maximum coincidence multiplicity for each of
the bands analyzed. The maximum multiplicity is an integer that varies from 3
to 20 because we require coincidence multiplicity of at least 3, 
and 20 is the number of time frames analyzed. 

The bottom panel of Fig.~\ref{fig:tdfstat_O1_Coinc_FA} shows the results for
the false alarm probability of coincidence for the coincidence with the maximum
multiplicity. This false alarm probability is calculated using the formula from the
Appendix of \cite{VSR1TDFstat}.

We define outliers as those coincidences with false alarm probabilities less than $0.1$\%
This criterion was adopted in our Virgo data 
search~\cite{VSR1TDFstat} and also in one of \EatH searches
~\cite{S4IncoherentPaper}. 
From the analysis we have excluded bands highly perturbed by 
violin modes and their harmonics. Thus the following four bands 
were vetoed: $[500, 509]$ Hz,  $[1001, 1025]$ Hz,  $[1483, 1511]$ Hz, and $[1957, 1966]$ Hz.
As a result we obtained 74 outliers.
The parameters of these outliers are listed in Table
\ref{tab:TDFstat_outliers}. 
The parameters of a given coincidence are calculated as the mean values 
of the parameters of the candidates that enter a given coincidence. Among the 74
outliers, 10 are identified with the hardware injections. 
Table \ref{tab:tdfstat_hi} presents the estimated parameters obtained 
for these hardware injections, along with 
the absolute errors of the reconstructed parameters
(the differences with respect to the injected parameters).
The remaining 64 outliers include 10 that are seen only in H1 data, 1 in only the L1 data. 
3 of the outliers are absent in the last one third of the data, 1 present in the first one third of the data, 
and 2 have a wandering frequency that increases in the first third of the run, is constant in the second third, 
decreases in the last one third of the run. 
The remaining 47 outliers seem to be harmonics of the same interference in the data. 
The distribution of the {\Fstat} in a given
time frame has approximately the same morphology for all the harmonics.
The outliers are present both in H1 and L1 but not always in coincidence.
When they are present in both detectors their SNRs are not consistent, and are at times much louder in L1.
Moreover the outliers appear in the stretch of a
two day data segment where 87\% of data are zeros. The remaining data in 
that segment are mainly a noise free modulated periodic signal.
We conclude that the interference originates from the detectors themselves 
as it clearly appears in a stretch of data with a small fraction of science data.
Consequently no credible gravitational wave candidates were found.

\section{Conclusions}

We have performed the most sensitive all-sky search to date for continuous gravitational waves 
in the range 475-2000\,Hz using three different methods. We explored both positive and negative spindowns and placed upper limits on expected and unexpected sources.  
Figure \ref{fig:powerflux_O1_upper_limits} shows a summary of the strain amplitude upper limits obtained for the three pipelines.  One pipeline ({\em PowerFlux}) presents strict all-sky limits for circular-polarization and linear polarisation sources. The other two  pipelines ({\em SkyHough} and  \td\ ) present  frequentist population-averaged limits over the full sky and source polarisation.

At the highest frequencies we are sensitive to neutron stars with an equatorial 
ellipticity as small as $\sci{1.8}{-7}$ and as far away as $1$\,kpc for favorable spin orientations.
The maximum ellipticity a neutron star can theoretically support is at least $\sci{1}{-5}$ according to \cite{crust_limit, crust_limit2}. Our results exclude such maximally deformed pulsars above a $200$\,Hz stellar rotation frequency ($400$\,Hz gravitational frequency) within $1$\,kpc.

Outliers from the initial stages of each search method were meticulously followed up, but no candidates from any search survived scrutiny.  

The use of the {\em Universal} statistic and {\em Loosely Coherent} algorithms allowed us to establish upper limits and achieve good detection efficiency (relative to the upper limit) in all frequency ranges, including highly contaminated areas. 

{\em SkyHough} pipeline added a viewpoint of robust Hough algorithm. Although the decrease in the sky grid
resolution at 1200 Hz, tuned to reduce computational load, produced a jump in sensitivity of about 20 \%,
this method offers an independent  check of  the other results. Future searches will use longer SFT time duration 
 to  allow the attainment of sensitivity close to {\em PowerFlux} at a reduced computational cost.

The use of a shorter coherent time and a looser grid for \td\ pipeline in the high frequency search with respect to the low frequency search resulted in loss of sensitivity by a factor of 3. With an increasing available computing power the search of the next data set will be performed with a considerably longer coherent time that should results in a sensitivity slightly better than the worse case for the PowerFlux analysis.

\section*{Acknowledgments}


The authors gratefully acknowledge the support of the United States
National Science Foundation (NSF) for the construction and operation of the
LIGO Laboratory and Advanced LIGO as well as the Science and Technology Facilities Council (STFC) of the
United Kingdom, the Max-Planck-Society (MPS), and the State of
Niedersachsen/Germany for support of the construction of Advanced LIGO 
and construction and operation of the GEO600 detector. 
Additional support for Advanced LIGO was provided by the Australian Research Council.
The authors gratefully acknowledge the Italian Istituto Nazionale di Fisica Nucleare (INFN),  
the French Centre National de la Recherche Scientifique (CNRS) and
the Foundation for Fundamental Research on Matter supported by the Netherlands Organisation for Scientific Research, 
for the construction and operation of the Virgo detector
and the creation and support  of the EGO consortium. 
The authors also gratefully acknowledge research support from these agencies as well as by 
the Council of Scientific and Industrial Research of India, 
the Department of Science and Technology, India,
the Science \& Engineering Research Board (SERB), India,
the Ministry of Human Resource Development, India,
the Spanish  Agencia Estatal de Investigaci\'on,
the Vicepresid\`encia i Conselleria d'Innovaci\'o, Recerca i Turisme and the Conselleria d'Educaci\'o i Universitat del Govern de les Illes Balears,
the Conselleria d'Educaci\'o, Investigaci\'o, Cultura i Esport de la Generalitat Valenciana,
the National Science Centre of Poland,
the Swiss National Science Foundation (SNSF),
the Russian Foundation for Basic Research, 
the Russian Science Foundation,
the European Commission,
the European Regional Development Funds (ERDF),
the Royal Society, 
the Scottish Funding Council, 
the Scottish Universities Physics Alliance, 
the Hungarian Scientific Research Fund (OTKA),
the Lyon Institute of Origins (LIO),
the Paris \^{I}le-de-France Region, 
the National Research, Development and Innovation Office Hungary (NKFI), 
the National Research Foundation of Korea,
Industry Canada and the Province of Ontario through the Ministry of Economic Development and Innovation, 
the Natural Science and Engineering Research Council Canada,
the Canadian Institute for Advanced Research,
the Brazilian Ministry of Science, Technology, Innovations, and Communications,
the International Center for Theoretical Physics South American Institute for Fundamental Research (ICTP-SAIFR), 
the Research Grants Council of Hong Kong,
the National Natural Science Foundation of China (NSFC),
the Leverhulme Trust, 
the Research Corporation, 
the Ministry of Science and Technology (MOST), Taiwan
and
the Kavli Foundation.
The authors gratefully acknowledge the support of the NSF, STFC, MPS, INFN, CNRS, PL-Grid and the
State of Niedersachsen/Germany for provision of computational resources.

This document has been assigned LIGO Laboratory document number \texttt{LIGO-P1700164-v18}.

\appendix
\section{Outlier tables}

{\em PowerFlux} outliers passing all stages of automated followup from 475-2000\,Hz band are separated into four tables. Table \ref{tab:PowerFluxOutliers} shows all outliers spanning 2 or more segments and outside heavily contaminated frequency range 1000-1033\,Hz. Table \ref{tab:PowerFluxOutliersViolin} shows outliers inside the contaminated region 1000-1033\,Hz. Lastly tables \ref{tab:PowerFluxOutliersShortMid} and \ref{tab:PowerFluxOutliersShortHigh} show ``short'' outliers using only 1 segment (approximately a month) of data. Table \ref{tab:PowerFluxOutliersShortMid} shows such short outliers below 1100\,Hz, while table \ref{tab:PowerFluxOutliersShortHigh} lists short outliers above 1100\,Hz. The splitting frequency of 1100\,Hz was chosen only to put similar numbers of outliers in each table.

\begin{table*}[htbp]
\begin{center}
\scriptsize
\begin{tabular}{D{.}{.}{2}D{.}{.}{2}cD{.}{.}{5}D{.}{.}{4}D{.}{.}{4}D{.}{.}{4}l}\hline
\multicolumn{1}{c}{Idx} & \multicolumn{1}{c}{SNR}  & \multicolumn{1}{c}{Segment} &  \multicolumn{1}{c}{Frequency} & \multicolumn{1}{c}{Spindown} &  \multicolumn{1}{c}{$\RAJ$}  & \multicolumn{1}{c}{$\DECJ$} & Description \\
\multicolumn{1}{c}{}	&  \multicolumn{1}{c}{}	&  \multicolumn{1}{c}{}	& \multicolumn{1}{c}{Hz}	&  \multicolumn{1}{c}{nHz/s} & \multicolumn{1}{c}{degrees} & \multicolumn{1}{c}{degrees} & \\
\hline \hline
3 & 3886 & $[0,2]$ & 1220.55536 & -0.300 & 229.053 & -2.107 & Injection 7, very different H1 and L1 sensitivities\\
4 & 456 & $[1,2]$ & 848.97002 & -0.350 & 37.141 & -29.612 & Injection 1, L1 much more sensitive than H1\\
5 & 375 & $[1,2]$ & 763.84713 & 0.000 & 198.171 & 75.664 & Injection 9, loud enough to be visible in background of H1 and L1.\\
6 & 286 & $[0,2]$ & 575.16361 & 0.000 & 215.370 & 3.558 & Injection 2, L1 is more sensitive than H1\\
14 & 126 & $[0,1]$ & 1080.00097 & 0.200 & 271.159 & 66.681 & Exceptionally strong coincident bin-centered lines at 1080 Hz.\\
16 & 85 & $[0,2]$ & 1487.98795 & -9.550 & 144.132 & -66.819 & Strong bin-centered line in H1 at 1488.00 Hz\\
19 & 68 & $[0,2]$ & 1220.43752 & -1.975 & 169.199 & -0.960 & Induced by injection 7.\\
24 & 41 & $[1,2]$ & 767.96349 & 1.475 & 118.599 & 78.067 & Strong bin-centered line in H1 at 768 Hz\\
25 & 37 & $[1,2]$ & 615.00752 & -4.700 & 202.130 & 63.562 & Strong broad line in L1\\
26 & 37 & $[0,1]$ & 713.38012 & -3.900 & 223.547 & 64.304 & Strong bin-centered line in L1 at 713.400 Hz\\
27 & 36 & $[1,2]$ & 585.38340 & -9.550 & 207.405 & 0.724 & Strong bin-centered line in L1 at 585.400 Hz\\
31 & 29 & $[0,2]$ & 1220.46981 & -7.650 & 177.333 & 53.647 & Induced by injection 7.\\
32 & 29 & $[0,1]$ & 943.98085 & 1.250 & 341.520 & 70.413 & Strong bin-centered line in H1 at 944.00 Hz\\
33 & 28 & $[0,2]$ & 910.06257 & 1.475 & 100.432 & 80.276 & Strong broad line in H1\\
34 & 27 & $[0,1]$ & 980.41316 & 0.425 & 68.498 & 19.939 & Strong bin-centered line in L1 at 980.500 Hz, line in H1\\
35 & 26 & $[0,1]$ & 1457.59127 & 0.900 & 5.161 & 22.809 & Highly non-stationary L1 data \\
36 & 26 & $[0,2]$ & 767.97611 & -4.025 & 106.321 & -57.243 & Bin-centered line in H1 at 768.00 Hz \\
37 & 26 & $[0,2]$ & 1255.99635 & -1.725 & 100.106 & -67.630 & Line in H1 at 1256 Hz \\
40 & 23 & $[1,2]$ & 1456.03964 & -1.600 & 215.391 & -69.386 & Highly non-stationary H1 data, line at 1456.00 Hz\\
41 & 23 & $[0,1]$ & 2000.00108 & -4.275 & 146.821 & -64.950 & Line in H1, violin mode harmonic region\\
42 & 23 & $[0,1]$ & 831.94019 & -7.550 & 139.056 & -28.186 & Bin-centered line in H1 at 832.00 Hz\\
43 & 22 & $[0,1]$ & 918.82255 & -1.525 & 294.016 & -66.661 & Strong broad line in L1\\
44 & 21 & $[0,2]$ & 899.29679 & 1.475 & 298.627 & 26.700 & Strong broad line in H1\\
45 & 21 & $[1,2]$ & 968.29014 & -7.550 & 105.510 & -69.138 & Mismatch in SNR between H1 and L1\\
46 & 21 & $[0,1]$ & 943.94642 & -6.450 & 72.434 & -43.175 & Bin-centered line in H1 at 944.00 Hz\\
47 & 20 & $[0,2]$ & 1167.94911 & 1.325 & 81.001 & -36.869 & Bin-centered line in H1 at 1168.00 Hz\\
48 & 20 & $[0,1]$ & 1983.05344 & -4.350 & 28.614 & -29.172 & Line in L1 at 1983.0994 Hz\\
49 & 20 & $[0,2]$ & 1393.47837 & -0.075 & 269.418 & -38.074 & Appears to be associated with injection 4\\
50 & 20 & $[1,2]$ & 559.75418 & -4.975 & 99.663 & 2.943 & Bin-centered line in L1 at 559.800 Hz\\
51 & 18 & $[0,1]$ & 1471.00891 & -0.600 & 19.261 & 82.386 & Highly non-stationary H1 spectrum\\
52 & 18 & $[1,2]$ & 629.87432 & -4.025 & 208.388 & 61.733 & Strong broad line in L1\\
53 & 17 & $[0,1]$ & 918.73177 & 1.325 & 77.766 & -40.562 & Strong broad line in L1\\
54 & 17 & $[1,2]$ & 623.96957 & -0.075 & 198.261 & 63.320 & Bin-centered line in H1 at 624.00 Hz\\
55 & 17 & $[1,2]$ & 588.29660 & -2.325 & 20.174 & 63.144 & Strong bin-centered line in L1 at 588.300 Hz\\
57 & 17 & $[0,1]$ & 1455.93002 & -1.075 & 72.852 & -37.095 & Very non-stationary H1 spectrum, line at 1456.00 Hz\\
58 & 17 & $[0,1]$ & 567.99073 & 0.525 & 275.690 & 77.944 & Strange coincident lines at 568.00 Hz\\
59 & 16 & $[1,2]$ & 906.51613 & -5.650 & 114.204 & 7.807 & Bin-centered line in L1 at 906.600 Hz\\
60 & 16 & $[0,2]$ & 588.31398 & -5.550 & 208.182 & -49.133 & Strong bin-centered line in L1 at 588.300 Hz\\
62 & 16 & $[0,1]$ & 1400.00418 & 0.675 & 85.821 & -67.453 & Bin-centered line in H1 at 1400.00 Hz\\
63 & 16 & $[0,2]$ & 575.09743 & -10.525 & 223.480 & 54.438 & Induced by injection 2\\
64 & 15 & $[1,2]$ & 1055.67464 & -8.850 & 52.210 & -62.000 & Poor coherence between H1 and L1\\
65 & 15 & $[0,2]$ & 918.75333 & -4.250 & 259.272 & 65.613 & Strong broad line in L1\\
66 & 14 & $[0,1]$ & 600.00424 & -5.950 & 194.962 & -83.060 & Strong line in H1 near 600 Hz\\
67 & 14 & $[0,1]$ & 906.72776 & -4.475 & 95.914 & 8.234 & Strong broad line in H1\\
68 & 13 & $[1,2]$ & 1198.55097 & 1.175 & 197.933 & 80.202 & Strong broad line in H1\\
69 & 13 & $[0,2]$ & 627.89160 & -8.200 & 225.017 & 32.253 & Bin-centered line in L1 at 627.900 Hz\\
71 & 12 & $[1,2]$ & 966.05168 & -5.725 & 290.296 & 45.961 & H1 and L1 SNR inconsistent\\
72 & 12 & $[0,1]$ & 956.52184 & -5.950 & 96.516 & 6.398 & \\

\hline
\end{tabular}
\caption[Outliers that passed detection pipeline]{Outliers that passed the PowerFlux detection pipeline spanning more than one segment and excluding the 1000-1033\,Hz region heavily contaminated with violin modes. Only the highest-SNR outlier is shown for each 0.1\,Hz frequency region. Outliers marked with ``line'' had strong narrowband disturbances identified near the outlier location. Outliers marked as ``non Gaussian'' were identified as having non Gaussian statistics in their power sums, often due to a very steeply sloping spectrum.  Segment column reports the set of
contiguous segments of the data that produced the outlier, as
described in \ref{sec:results}.
Frequencies are converted to epoch GPS $1130529362$.}
\label{tab:PowerFluxOutliers}
\end{center}
\end{table*}

\begin{table*}[htbp]
\begin{center}
\scriptsize
\begin{tabular}{D{.}{.}{2}D{.}{.}{2}cD{.}{.}{5}D{.}{.}{4}D{.}{.}{4}D{.}{.}{4}l}\hline
\multicolumn{1}{c}{Idx} & \multicolumn{1}{c}{SNR}  & \multicolumn{1}{c}{Segment} &  \multicolumn{1}{c}{Frequency} & \multicolumn{1}{c}{Spindown} &  \multicolumn{1}{c}{$\RAJ$}  & \multicolumn{1}{c}{$\DECJ$} & \\
\multicolumn{1}{c}{}	&  \multicolumn{1}{c}{}	&  \multicolumn{1}{c}{}	& \multicolumn{1}{c}{Hz}	&  \multicolumn{1}{c}{nHz/s} & \multicolumn{1}{c}{degrees} & \multicolumn{1}{c}{degrees} & \\
\hline \hline
73 & 122634 & $[0,0]$ & 998.67165 & -6.050 & 34.496 & -58.000 & \\
74 & 76138 & $[0,0]$ & 998.61134 & 1.175 & 50.986 & 18.219 & \\
78 & 485 & $[2,2]$ & 512.01668 & -3.425 & 22.826 & -88.770 & \\
83 & 69 & $[1,1]$ & 832.01071 & -6.600 & 178.258 & -75.767 & \\
84 & 61 & $[1,1]$ & 863.96498 & -7.225 & 207.792 & 54.356 & \\
86 & 52 & $[1,1]$ & 952.02462 & -4.200 & 156.420 & -86.793 & \\
87 & 48 & $[1,1]$ & 781.48875 & -9.175 & 227.909 & 39.730 & \\
89 & 44 & $[1,1]$ & 1079.93838 & -6.050 & 185.624 & 58.142 & \\
96 & 28 & $[0,0]$ & 1099.69279 & -9.500 & 62.525 & -17.371 & \\
97 & 28 & $[2,2]$ & 918.70042 & -5.975 & 135.889 & -27.388 & \\
102 & 25 & $[0,0]$ & 945.25946 & -4.425 & 105.182 & -2.896 & \\
108 & 20 & $[1,1]$ & 568.53389 & -5.075 & 270.104 & -61.403 & \\
109 & 20 & $[2,2]$ & 1080.11043 & -7.325 & 307.143 & -1.254 & \\
110 & 19 & $[2,2]$ & 824.02132 & -5.700 & 147.900 & -86.736 & \\
111 & 19 & $[0,0]$ & 899.25908 & -5.650 & 337.827 & -21.074 & \\
113 & 19 & $[2,2]$ & 990.04856 & 0.300 & 135.434 & -26.528 & \\
114 & 19 & $[2,2]$ & 716.23123 & -0.075 & 168.445 & 20.834 & \\
115 & 19 & $[1,1]$ & 568.01764 & -2.225 & 251.325 & -89.632 & \\
116 & 19 & $[1,1]$ & 1096.02101 & -4.450 & 133.692 & -83.005 & \\
117 & 19 & $[0,0]$ & 922.55918 & -0.450 & 64.475 & 4.328 & \\
119 & 18 & $[2,2]$ & 1088.01257 & -10.825 & 248.325 & 39.022 & \\
121 & 18 & $[2,2]$ & 900.87618 & -4.750 & 308.565 & 23.094 & \\
122 & 18 & $[2,2]$ & 900.73436 & -5.900 & 161.021 & -19.505 & \\
123 & 17 & $[2,2]$ & 523.61892 & -6.825 & 240.077 & -55.972 & \\
124 & 17 & $[1,1]$ & 475.32726 & -6.025 & 207.149 & 78.036 & \\
126 & 17 & $[1,1]$ & 1088.04594 & 0.325 & 18.340 & -52.698 & \\
129 & 17 & $[1,1]$ & 1095.98516 & -10.525 & 159.987 & -62.138 & \\
130 & 17 & $[1,1]$ & 475.36243 & -8.625 & 283.160 & -83.890 & \\
131 & 16 & $[2,2]$ & 625.01993 & -5.925 & 333.353 & 50.108 & \\
132 & 16 & $[0,0]$ & 912.06903 & -2.325 & 281.739 & -53.318 & \\
133 & 16 & $[2,2]$ & 716.37292 & -8.400 & 306.163 & 12.283 & \\
134 & 16 & $[2,2]$ & 1091.97016 & -5.475 & 257.005 & -45.295 & \\
135 & 16 & $[0,0]$ & 922.66069 & 1.400 & 4.635 & -37.224 & \\
137 & 16 & $[1,1]$ & 1085.88189 & -3.275 & 222.923 & 41.844 & \\
138 & 16 & $[1,1]$ & 799.61576 & -4.225 & 305.876 & 58.952 & \\
141 & 16 & $[1,1]$ & 945.43339 & -4.575 & 277.653 & -1.384 & \\
143 & 16 & $[0,0]$ & 1063.98385 & -0.450 & 89.302 & -58.822 & \\
144 & 16 & $[2,2]$ & 874.92611 & -5.900 & 198.168 & 36.620 & \\
147 & 16 & $[2,2]$ & 1080.26045 & -5.425 & 147.600 & -21.563 & \\
148 & 16 & $[2,2]$ & 991.13399 & -9.700 & 217.041 & 21.846 & \\
149 & 16 & $[1,1]$ & 920.03446 & -5.200 & 309.442 & -84.932 & \\
152 & 15 & $[0,0]$ & 943.20137 & -8.275 & 60.640 & -34.099 & \\
153 & 15 & $[1,1]$ & 971.53220 & -1.200 & 270.236 & 33.046 & \\
154 & 15 & $[2,2]$ & 900.74745 & -7.825 & 165.665 & -30.418 & \\
156 & 15 & $[1,1]$ & 945.41047 & -10.600 & 260.757 & 3.250 & \\
159 & 15 & $[2,2]$ & 700.07700 & -1.325 & 143.438 & 53.430 & \\
160 & 15 & $[1,1]$ & 961.40660 & -8.400 & 318.893 & 27.718 & \\
161 & 15 & $[1,1]$ & 1054.71444 & -9.800 & 0.704 & -4.956 & \\
165 & 14 & $[2,2]$ & 831.64901 & -3.125 & 194.537 & -39.518 & \\
173 & 14 & $[1,1]$ & 739.29278 & -0.600 & 318.296 & -43.429 & \\
176 & 14 & $[0,0]$ & 718.00248 & -5.675 & 213.134 & -49.747 & \\
178 & 14 & $[1,1]$ & 669.61556 & -1.100 & 57.094 & -34.323 & \\
179 & 14 & $[0,0]$ & 775.14530 & -8.450 & 244.756 & -52.288 & \\
181 & 14 & $[0,0]$ & 1039.11823 & -1.375 & 313.591 & 35.538 & \\
182 & 14 & $[2,2]$ & 754.30629 & -5.450 & 16.599 & 47.778 & \\
183 & 14 & $[1,1]$ & 633.73616 & -10.900 & 11.121 & -54.400 & \\
192 & 13 & $[1,1]$ & 1069.18221 & -4.850 & 136.156 & -16.451 & \\
196 & 13 & $[1,1]$ & 583.96498 & -9.500 & 311.580 & 41.127 & \\
206 & 13 & $[2,2]$ & 758.50361 & -2.400 & 136.803 & -35.273 & \\
207 & 13 & $[0,0]$ & 1087.96981 & -9.250 & 54.309 & -60.667 & \\
209 & 12 & $[2,2]$ & 662.79818 & -5.200 & 219.379 & 35.883 & \\
211 & 12 & $[0,0]$ & 895.31856 & -9.575 & 242.924 & 15.227 & \\

\hline
\end{tabular}
\caption[Outliers that passed detection pipeline]{Outliers below 1100\,Hz that passed the PowerFlux detection pipeline spanning only one segment, excluding  1000-1033\,Hz region heavily contaminated with violin modes. Only the highest-SNR outlier is shown for each 0.1\,Hz frequency region.  Segment column reports the set of
contiguous segments of the data that produced the outlier, as
described in \ref{sec:results}.
Frequencies are converted to epoch GPS $1130529362$.}
\label{tab:PowerFluxOutliersShortMid}
\end{center}
\end{table*}

\begin{table*}[htbp]
\begin{center}
\scriptsize
\begin{tabular}{D{.}{.}{2}D{.}{.}{2}cD{.}{.}{5}D{.}{.}{4}D{.}{.}{4}D{.}{.}{4}l}\hline
\multicolumn{1}{c}{Idx} & \multicolumn{1}{c}{SNR}  & \multicolumn{1}{c}{Segment} &  \multicolumn{1}{c}{Frequency} & \multicolumn{1}{c}{Spindown} &  \multicolumn{1}{c}{$\RAJ$}  & \multicolumn{1}{c}{$\DECJ$} & \\
\multicolumn{1}{c}{}	&  \multicolumn{1}{c}{}	&  \multicolumn{1}{c}{}	& \multicolumn{1}{c}{Hz}	&  \multicolumn{1}{c}{nHz/s} & \multicolumn{1}{c}{degrees} & \multicolumn{1}{c}{degrees} & \\
\hline \hline
75 & 5854 & $[0,0]$ & 1456.14766 & -0.175 & 136.769 & -41.785 & \\
76 & 2713 & $[1,1]$ & 1987.38812 & -8.100 & 115.653 & -70.974 & \\
80 & 105 & $[1,1]$ & 1824.00927 & -8.250 & 126.515 & -75.314 & \\
81 & 91 & $[2,2]$ & 1393.56417 & -10.075 & 318.820 & -10.426 & \\
82 & 72 & $[0,0]$ & 1327.89729 & -7.325 & 140.109 & 69.425 & \\
85 & 59 & $[1,1]$ & 1872.06302 & -7.600 & 348.724 & -87.309 & \\
88 & 44 & $[0,0]$ & 1135.96045 & -7.525 & 218.121 & 64.723 & \\
90 & 41 & $[1,1]$ & 1997.31629 & -6.175 & 61.621 & -65.686 & \\
91 & 37 & $[0,0]$ & 1369.76707 & -0.050 & 187.539 & 59.193 & \\
92 & 36 & $[0,0]$ & 1999.90597 & 0.750 & 20.310 & 70.095 & \\
95 & 31 & $[0,0]$ & 1690.86031 & -2.875 & 96.307 & -14.940 & \\
98 & 27 & $[0,0]$ & 1999.78615 & -5.300 & 81.182 & 32.271 & \\
99 & 27 & $[1,1]$ & 1247.54194 & -6.900 & 128.268 & -45.602 & \\
103 & 25 & $[0,0]$ & 1999.83424 & -0.450 & 61.487 & 47.583 & \\
104 & 23 & $[1,1]$ & 1446.70535 & -7.725 & 22.583 & -68.576 & \\
105 & 21 & $[2,2]$ & 1393.29262 & -8.600 & 168.938 & 24.362 & \\
106 & 21 & $[0,0]$ & 1372.62964 & -3.400 & 89.257 & -41.198 & \\
107 & 20 & $[0,0]$ & 1135.90451 & -9.050 & 118.203 & -25.813 & \\
112 & 19 & $[2,2]$ & 1393.44556 & -10.300 & 179.179 & -86.352 & \\
120 & 18 & $[1,1]$ & 1262.53007 & -1.025 & 132.290 & -51.838 & \\
125 & 17 & $[1,1]$ & 1213.68816 & -7.975 & 347.921 & 41.708 & \\
127 & 17 & $[1,1]$ & 1290.46538 & -2.250 & 56.528 & -56.177 & \\
128 & 17 & $[1,1]$ & 1463.16241 & -3.050 & 34.883 & 37.388 & \\
136 & 16 & $[1,1]$ & 1424.20719 & -10.250 & 143.258 & 54.532 & \\
139 & 16 & $[1,1]$ & 1335.54724 & 0.000 & 27.733 & -76.368 & \\
140 & 16 & $[0,0]$ & 1213.56733 & -4.925 & 104.723 & 66.604 & \\
142 & 16 & $[1,1]$ & 1276.81304 & -7.075 & 58.235 & -29.473 & \\
145 & 16 & $[0,0]$ & 1907.05681 & -5.250 & 272.503 & -46.509 & \\
146 & 16 & $[2,2]$ & 1528.32712 & -4.950 & 37.441 & -60.096 & \\
150 & 15 & $[1,1]$ & 1459.94901 & -9.950 & 163.428 & -22.664 & \\
151 & 15 & $[2,2]$ & 1401.51757 & -7.250 & 355.062 & -38.577 & \\
155 & 15 & $[1,1]$ & 1138.62182 & -4.250 & 90.540 & -33.848 & \\
157 & 15 & $[1,1]$ & 1256.01957 & -9.275 & 176.253 & -78.174 & \\
158 & 15 & $[1,1]$ & 1211.07792 & -0.475 & 348.328 & 79.398 & \\
162 & 14 & $[1,1]$ & 1463.20594 & -5.200 & 15.555 & 28.395 & \\
163 & 14 & $[0,0]$ & 1219.64849 & 0.050 & 179.645 & -29.748 & \\
164 & 14 & $[2,2]$ & 1264.11240 & -5.100 & 313.522 & 18.441 & \\
166 & 14 & $[2,2]$ & 1295.86238 & -3.900 & 181.282 & -55.826 & \\
167 & 14 & $[2,2]$ & 1395.17951 & 0.350 & 162.187 & -56.606 & \\
168 & 14 & $[0,0]$ & 1264.56939 & -3.875 & 260.820 & 23.151 & \\
169 & 14 & $[2,2]$ & 1288.87309 & -1.900 & 337.759 & -17.706 & \\
170 & 14 & $[2,2]$ & 1203.61330 & -3.825 & 170.297 & -17.434 & \\
171 & 14 & $[1,1]$ & 1368.72368 & -2.875 & 153.808 & -53.399 & \\
172 & 14 & $[1,1]$ & 1337.95543 & -3.150 & 92.898 & -48.705 & \\
174 & 14 & $[0,0]$ & 1405.16819 & -2.050 & 358.179 & -32.990 & \\
175 & 14 & $[1,1]$ & 1230.31634 & 0.800 & 205.853 & 25.811 & \\
177 & 14 & $[0,0]$ & 1352.50502 & -8.400 & 294.685 & -2.192 & \\
180 & 14 & $[2,2]$ & 1421.97637 & -1.675 & 216.086 & 77.049 & \\
184 & 14 & $[1,1]$ & 1384.01262 & -0.250 & 344.524 & -69.144 & \\
185 & 14 & $[0,0]$ & 1251.71109 & -0.375 & 88.753 & 47.800 & \\
186 & 14 & $[1,1]$ & 1180.70083 & -0.575 & 74.681 & -30.772 & \\
187 & 14 & $[2,2]$ & 1404.40200 & -6.400 & 105.584 & 46.302 & \\
188 & 13 & $[0,0]$ & 1329.97102 & -4.750 & 175.844 & 46.741 & \\
189 & 13 & $[2,2]$ & 1130.57326 & -4.000 & 103.921 & 43.754 & \\
190 & 13 & $[1,1]$ & 1302.48986 & -5.450 & 186.382 & -59.427 & \\
191 & 13 & $[2,2]$ & 1248.31576 & 0.875 & 85.718 & -10.040 & \\
193 & 13 & $[1,1]$ & 1107.06549 & -6.450 & 127.799 & -10.620 & \\
194 & 13 & $[2,2]$ & 1451.70229 & -9.875 & 238.959 & 46.016 & \\
195 & 13 & $[1,1]$ & 1296.76012 & -4.750 & 260.379 & 33.317 & \\
197 & 13 & $[0,0]$ & 1171.26882 & -4.150 & 15.889 & -37.954 & \\
198 & 13 & $[0,0]$ & 1165.20479 & -4.825 & 60.686 & -23.324 & \\
199 & 13 & $[1,1]$ & 1164.53396 & -6.750 & 284.184 & 31.085 & \\
200 & 13 & $[1,1]$ & 1113.03840 & -8.575 & 194.191 & -63.810 & \\
201 & 13 & $[1,1]$ & 1266.40557 & -9.275 & 359.080 & 18.866 & \\
202 & 13 & $[1,1]$ & 1177.23828 & -6.950 & 302.299 & 65.853 & \\
203 & 13 & $[1,1]$ & 1285.67481 & -2.750 & 346.041 & -33.719 & \\
204 & 13 & $[1,1]$ & 1186.91571 & -8.950 & 211.272 & 18.279 & \\
205 & 13 & $[0,0]$ & 1432.28413 & -10.450 & 55.297 & -35.353 & \\
208 & 13 & $[0,0]$ & 1132.56792 & -6.625 & 248.673 & 37.290 & \\
210 & 12 & $[2,2]$ & 1257.08005 & -0.850 & 117.394 & -38.201 & \\
212 & 12 & $[1,1]$ & 1321.09437 & -4.050 & 67.216 & -35.597 & \\
213 & 12 & $[1,1]$ & 1324.20852 & -7.150 & 104.807 & 56.301 & \\

\hline
\end{tabular}
\caption[Outliers that passed detection pipeline]{Outliers above 1100\,Hz that passed the PowerFlux detection pipeline spanning only one segment. Only the highest-SNR outlier is shown for each 0.1\,Hz frequency region.  Segment column reports the set of
contiguous segments of the data that produced the outlier, as
described in \ref{sec:results}.
Frequencies are converted to epoch GPS $1130529362$.}
\label{tab:PowerFluxOutliersShortHigh}
\end{center}
\end{table*}

\begin{table*}[htbp]
\begin{center}
\scriptsize
\begin{tabular}{D{.}{.}{2}D{.}{.}{2}cD{.}{.}{5}D{.}{.}{4}D{.}{.}{4}D{.}{.}{4}l}\hline
\multicolumn{1}{c}{Idx} & \multicolumn{1}{c}{SNR}  & \multicolumn{1}{c}{Segment} &  \multicolumn{1}{c}{Frequency} & \multicolumn{1}{c}{Spindown} &  \multicolumn{1}{c}{$\RAJ$}  & \multicolumn{1}{c}{$\DECJ$}  \\
\multicolumn{1}{c}{}	&  \multicolumn{1}{c}{}	&  \multicolumn{1}{c}{}	& \multicolumn{1}{c}{Hz}	&  \multicolumn{1}{c}{nHz/s} & \multicolumn{1}{c}{degrees} & \multicolumn{1}{c}{degrees} \\
\hline \hline
1 & 20746 & $[1,2]$ & 1019.64700 & -4.625 & 246.424 & 80.922 & Extremely strong bin-centered line in L1\\
2 & 20438 & $[0,1]$ & 1020.36752 & -1.750 & 253.492 & 63.937 & Lines in H1 and L1\\
7 & 283 & $[0,2]$ & 1008.00325 & -10.825 & 221.934 & 43.985 & Very strong line in L1\\
8 & 264 & $[1,2]$ & 1008.12309 & -8.600 & 301.450 & -25.837 & Very strong line in L1\\
9 & 257 & $[0,1]$ & 1007.92946 & 0.575 & 90.115 & 11.329 & Very strong line in L1, line in H1 at different frequency\\
10 & 249 & $[1,2]$ & 1026.85819 & -9.925 & 169.924 & -66.143 & Forest of strong lines in L1\\
11 & 185 & $[1,2]$ & 1023.86681 & -5.350 & 314.269 & -4.805 & Forest of strong lines in L1\\
12 & 182 & $[1,2]$ & 1023.91746 & 1.250 & 153.699 & 75.645 & Extremely strong line in L1\\
13 & 133 & $[0,2]$ & 1012.64960 & -6.950 & 279.830 & -18.978 & Strong lines in L1, highly non-stationary spectrum. Disturbed H1 spectrum.\\
15 & 118 & $[1,2]$ & 1023.88441 & -5.925 & 164.783 & 12.160 & Forest of strong lines in L1\\
17 & 74 & $[0,2]$ & 1032.24361 & -10.675 & 150.097 & -53.175 & Forest of strong lines in L1\\
18 & 72 & $[0,2]$ & 1026.77755 & -10.250 & 104.093 & -14.244 & Forest of strong lines in L1\\
20 & 59 & $[0,1]$ & 1032.80017 & -7.075 & 353.600 & -66.276 & Forest of strong lines in L1\\
21 & 53 & $[0,2]$ & 1031.18496 & -9.875 & 157.882 & -34.654 & Forest of strong lines in L1\\
22 & 50 & $[0,2]$ & 1026.93116 & -6.100 & 300.750 & 26.695 & Forest of strong lines in L1\\
23 & 43 & $[0,2]$ & 1030.85351 & -3.375 & 145.333 & 77.333 & Forest of strong lines in L1\\
28 & 36 & $[0,2]$ & 1029.16420 & -5.900 & 94.435 & -68.285 & Forest of strong lines in L1\\
29 & 32 & $[0,1]$ & 1006.53372 & -5.550 & 212.656 & -74.205 & Strange broad line in H1\\
30 & 31 & $[0,2]$ & 1032.22826 & -7.775 & 132.317 & -45.682 & Forest of strong lines in L1\\
38 & 24 & $[1,2]$ & 1026.10892 & -2.450 & 29.852 & -82.280 & Forest of strong lines in L1\\
39 & 24 & $[1,2]$ & 1026.06630 & -1.925 & 124.580 & -66.716 & Forest of strong lines in L1\\
56 & 17 & $[1,2]$ & 1016.00465 & -4.900 & 107.871 & 4.395 & Highly non-stationary L1 data\\
61 & 16 & $[0,1]$ & 1003.61312 & -1.175 & 108.017 & -37.989 & Strong broad line in H1\\
70 & 13 & $[1,2]$ & 1006.00859 & -6.325 & 112.936 & 5.218 & Bin-centered line in L1 at 1006.100 Hz, broad line in H1\\
77 & 510 & $[0,0]$ & 1027.01297 & 1.025 & 26.276 & 70.439 & \\
79 & 185 & $[1,1]$ & 1022.43734 & -2.425 & 117.977 & -56.277 & \\
93 & 36 & $[1,1]$ & 1027.31427 & 0.550 & 155.955 & 65.509 & \\
94 & 36 & $[0,0]$ & 1019.41689 & -9.050 & 310.849 & -53.911 & \\
100 & 27 & $[0,0]$ & 1006.51372 & -10.975 & 223.516 & 14.553 & \\
101 & 27 & $[0,0]$ & 1005.90983 & -4.925 & 270.705 & 72.119 & \\
118 & 18 & $[0,0]$ & 1000.00868 & -9.250 & 261.463 & 37.283 & \\

\hline
\end{tabular}
\caption[Outliers near violin modes that passed detection pipeline]{PowerFlux outliers in 1000-1033\,Hz region heavily contaminated with violin modes. Only the highest-SNR outlier is shown for each 0.1\,Hz frequency region. Outliers marked with ``line'' had strong narrowband disturbances identified near the outlier location. Outliers marked as ``non Gaussian'' were identified as having non Gaussian statistics in their power sums, often due to a very steeply sloping spectrum. Segment column reports the set of
contiguous segments of the data that produced the outlier, as
described in \ref{sec:results}.
Frequencies are converted to epoch GPS $1130529362$.}
\label{tab:PowerFluxOutliersViolin}
\end{center}
\end{table*}

Table~\ref{tab:sh_outliers} shows the parameters of the final 25 outliers
from the  {\em SkyHough} pipeline, along with comments on their likely origin. 
None of these outliers show evidence of being a credible gravitational wave signal.
\begin{table*}[htbp]
\begin{center}
\scriptsize
\begin{tabular}{ccccccccccccccccc}\hline
Idx  & Frequency  & $\alpha$ & $\delta$ & Spin-down & $s_{\textrm{mean}}$ & $\#_{\textrm{cluster}}$ & $\#_{\textrm{L1}}$ & $\#_{\textrm{H1}}$  & $s^*_{\textrm{L1}}$  & $s^*_{\textrm{H1}}$  & $s_{\textrm{max}}$  & Description \\
 &  [Hz] &  [rad] & [rad] & [nHz/s] &   &   &   &   &   &   &   & \\
\hline \hline \\
1 & 501.6000 & -1.4445 & 1.2596 & 0.9374 & 10.66 & 5 & 2 & 3 & 11.31 & 89.18 & 10.71 & Quad violin mode 1st harmonic region (H1 \& L1)\\
2 & 511.9968 & -1.4218 & 1.2070 & 0.6773 & 16.31 & 4927 & 298 & 226 & 10.47 & 101.36 & 18.73 & Quad violin mode 1st harmonic region (H1 \& L1)\\
3 & 512.0027 & 1.7085 & -1.1996 & -0.6071 & 16.33 & 3007 & 245 & 246 & 11.20 & 101.55 & 18.85 & Quad violin mode 1st harmonic region (H1 \& L1)\\
4 & 568.0011 & 1.5942 & -1.1783 & -0.1839 & 7.18 & 3867 & 415 & 125 & 8.82 & 9.81 & 9.05 & 8 Hz comb (H1 \& L1)\\
5 & 575.1635 & -2.5290 & 0.0526 & 0.0170 & 30.50 & 1974 & 275 & 78 & 46.66 & 26.54 & 33.75 & Hardware injection ip2\\
6 & 715.7250 & 1.0629 & -0.2049 & -2.0400 & 5.48 & 5 & 3 & 4 & 6.53 & 6.50 & 5.53 & Unknown\\
8 & 763.8507 & -2.7245 & 1.2888 & -0.5567 & 35.85 & 6064 & 297 & 91 & 41.29 & 43.43 & 42.33 & Hardware injection ip9\\
9 & 763.9016 & -2.1715 & 0.9109 & -7.1318 & 18.19 & 611 & 151 & 56 & 17.45 & 22.99 & 19.84 & Hardware injection child ip9\\
11 & 824.0035 & 1.6679 & -1.1996 & -0.7762 & 7.56 & 1111 & 81 & 123 & 8.09 & 10.83 & 8.43 & 8 Hz comb (H1 \& L1)\\
12 & 848.9657 & 0.6589 & -0.4414 & 0.5497 & 36.06 & 5329 & 342 & 117 & 48.63 & 37.64 & 42.17 & Hardware injection ip1\\
13 & 849.0020 & 0.4565 & -0.6807 & -4.0716 & 25.19 & 1983 & 331 & 108 & 31.08 & 29.57 & 29.35 & Hardware injection child ip1\\
14 & 895.9988 & -1.5481 & 1.1744 & 0.2368 & 10.33 & 244 & 35 & 79 & 6.48 & 69.62 & 11.45 & 8 Hz comb (H1 \& L1)\\
15 & 952.0018 & 1.5957 & -1.1797 & -0.3216 & 18.57 & 4353 & 355 & 189 & 18.36 & 27.59 & 21.86 & 8 Hz comb (H1 \& L1)\\
16 & 952.1017 & -0.3965 & -1.3294 & -9.8134 & 9.08 & 416 & 138 & 62 & 9.17 & 15.29 & 9.98 & 8 Hz comb (H1 \& L1)\\
17 & 1079.9981 & -1.5517 & 1.1798 & 0.3367 & 22.98 & 2639 & 402 & 129 & 51.28 & 17.88 & 25.90 & 8 Hz comb (H1 \& L1)\\
18 & 1080.0022 & 1.6073 & -1.1825 & -0.4562 & 22.95 & 5276 & 428 & 172 & 52.66 & 17.84 & 25.89 & 8 Hz comb (H1 \& L1)\\
19 & 1080.1007 & -0.2290 & -1.3906 & -9.9428 & 10.79 & 451 & 117 & 49 & 20.45 & 9.52 & 12.60 & 8 Hz comb (H1 \& L1)\\
21 & 1220.5492 & -2.2823 & 0.0725 & 0.5482 & 34.69 & 291 & 63 & 43 & 66.56 & 37.98 & 48.10 & Hardware injection ip7\\
22 & 1220.7094 & -1.6804 & -0.5910 & -9.6702 & 6.14 & 17 & 12 & 11 & 7.37 & 8.32 & 6.58 & Hardware injection child ip7\\
44 & 1475.0997 & 1.5636 & -1.1725 & -0.0308 & 10.87 & 42 & 8 & 19 & 6.64 & 77.42 & 11.72 & Quad violin mode 3rd harmonic region (H1 \& L1)\\
45 & 1482.5000 & -2.8976 & 1.0123 & 0.7317 & 9.04 & 2 & 1 & 2 & 6.58 & 51.78 & 9.05 & Quad violin mode 3rd harmonic region (H1 \& L1)\\
46 & 1487.8976 & 1.8780 & 1.1717 & -1.7738 & 6.69 & 2 & 1 & 2 & 6.53 & 10.19 & 6.75 & Quad violin mode 3rd harmonic region (H1)\\
66 & 1903.9302 & -1.8796 & 1.5402 & 0.1383 & 15.51 & 65 & 28 & 12 & 35.47 & 39.89 & 35.48 & 8 Hz comb (H1 \& L1)\\
67 & 1904.0020 & 1.5885 & -1.1737 & -0.4096 & 29.00 & 4779 & 340 & 141 & 34.94 & 40.65 & 36.82 & 8 Hz comb (H1 \& L1) \\
68 & 1904.1028 & 0.9560 & -1.3834 & -10.0406 & 15.11 & 925 & 194 & 51 & 16.36 & 24.82 & 19.12 & 8 Hz comb (H1 \& L1)\\
\hline
\end{tabular}
\caption[]{ {\em SkyHough} pipeline outliers in the range of frequencies between 475 and 2000 Hz after the population veto.
The table provides the frequency, spin-down and sky location of the cluster centers found by the {\em SkyHough}  search. 
$\#_{cluster}$ is the size of the cluster in terms of number of coincident pairs, 
$s_{max}$ and  $s_{mean}$ are the maximum and mean value of the cluster significance, 
$\#_{L1}$ and $\#_{H1}$ are the number of different candidates producing coincidence pairs from the different data sets, and $s^*_{L1}$ and $s^*_{H1}$ are the maximum significance values obtained by analysing the data from H1 and L1 separately.
Frequencies are converted to epoch GPS 1125972653.}
\label{tab:sh_outliers}
\end{center}
\end{table*}

Table~\ref{tab:TDFstat_outliers} presents the parameters of the final 74 outliers
from the {\td}  pipeline, along with comments on their likely causes.
None is a credible gravitational wave signal.

\begin{table*}[htbp]
\begin{center}
\scriptsize
\begin{tabular}{D{.}{.}{2}D{.}{.}{6}D{.}{.}{6}D{.}{.}{5}D{.}{.}{4}D{.}{.}{4}l}\hline
\multicolumn{1}{c}{Idx} & \multicolumn{1}{c}{FAP} & \multicolumn{1}{c}{Frequency} & \multicolumn{1}{c}{Spindown} &  \multicolumn{1}{c}{$\RAJ$}  & \multicolumn{1}{c}{$\DECJ$} & Description \\
\multicolumn{1}{c}{}	& \multicolumn{1}{c}{}	& \multicolumn{1}{c}{Hz}	&  \multicolumn{1}{c}{nHz/s} & \multicolumn{1}{c}{degrees} & \multicolumn{1}{c}{degrees} &  \\
\hline \hline
  1  & 9.1\times10^{-4} &  476.23802  &   -1.613  &  314.9096  &  -70.2754  &   harmonic of a detector interference         \\
  2  & 8.0\times10^{-4} &  486.89080  &   -0.061  &  304.1872  &   44.0319  &            harmonic of a detector interference         \\
  3  & 5.3\times10^{-4} &  487.61370  &   -1.133  &  268.2052  &   30.6643  &            absent in the last 1/3 of the data\\
  4  & 8.3\times10^{-4} &    492.22690  &   -0.615  &  280.4806  &   20.3164  &            harmonic of a detector interference         \\
  5  & 2.0\times10^{-5} &    499.26822  &    0.224  &  265.2608  &   70.6734  &            Present only in H1 \\
  6  & 3.3\times10^{-7} &    499.28018  &   -1.546  &  119.7586  &  -83.1059  &            Present only in H1 \\
  7  & 1.3\times10^{-4} &    518.14518  &   -0.251  &  320.7519  &   35.3491  &            harmonic of a detector interference         \\
  8  & 1.3\times10^{-4} &    531.94696  &    0.251  &  287.7129  &   29.1385  &            absent in the last 1/3 of the data\\
  9  & 1.3\times10^{-4} &    571.66195  &   -1.235  &  350.9658  &  -72.5879  &            harmonic of a detector interference         \\
 10  & 1.3\times10^{-4} &    575.16544  &    0.293  &  219.7073  &   12.2089  &            Injection 2\\
 11  & 1.3\times10^{-4} &    575.16377  &    0.016  &  203.8658  &  -27.1485  &            Injection 2\\
 12  &   5.9\times10^{-4} &  580.85725  &    0.104  &   31.4819  &  -66.8292  &            harmonic of a detector interference\\
 13  &   3.7\times10^{-4} &  593.93609  &   -0.340  &  195.5173  &  -85.6270  &            harmonic of a detector interference\\
 14  &   5.9\times10^{-4} &  603.61601  &   -2.460  &  253.8641  &   30.8522  &            harmonic of a detector interference\\
 15  &   2.7\times10^{-4} &  604.42590  &   -0.034  &  146.0965  &   25.3840  &            Present only in H1
\\
 16  &   2.8\times10^{-4} &  604.42583  &   -0.237  &  141.1892  &    4.7461  &            Present only in H1
\\
 17  &   3.9\times10^{-7} &  606.60486  &   -0.204  &  149.2635  &   26.6243  &            Present only in H1
\\
 18  &   1.3\times10^{-5} &  606.60513  &   -0.203  &  138.3114  &    1.3266  &            Present only in H1
\\
 19  &   5.4\times10^{-4} &  631.47115  &   -1.004  &  270.0083  &   43.2007  &            absent in the last 1/3 of the data\\
 20  &   1.9\times10^{-4} &  659.09677  &   -2.865  &  298.0274  &  -73.3156  &            harmonic of a detector interference\\
 21  &   9.9\times10^{-4} &  690.09526  &   -0.659  &  275.6423  &   32.2947  &            harmonic of a detector interference\\
 22  &   5.6\times10^{-4} &  735.36919  &   -0.679  &   66.2231  &  -82.7547  &            harmonic of a detector interference\\
 23  &   0 & 763.84721  &    0.050  &  197.8817  &   75.9108  &            Injection 9\\
 24  &   0 & 763.86856  &   -4.532  &  166.7853  &  -65.5177  &            Injection 9\\
 25  &   8.6\times10^{-4} &  769.53252  &   -2.470  &  329.3430  &  -80.4823  &            harmonic of a detector interference\\
 26  &   1.6\times10^{-4} &  787.45070  &   -0.803  &  298.4857  &   52.5868  &            harmonic of a detector interference\\
 27  &   1.2\times10^{-4} &  806.10968  &   -3.761  &  287.7636  &  -73.5434  &            harmonic of a detector interference\\
 28  &   5.0\times10^{-4} &  820.86500  &   -2.207  &  265.9691  &   40.4204  &            harmonic of a detector interference\\
 29  &   3.0\times10^{-4} &  820.86681  &   -0.189  &   48.1082  &  -81.1524  &            harmonic of a detector interference\\
 30  &   8.4\times10^{-4} &  831.52219  &   -0.389  &   52.4265  &  -81.1168  &            harmonic of a detector interference\\
 31  &   0 & 848.92226  &   -0.201  &  217.6862  &   26.1052  &            Injection 1\\
 32  &   0 & 848.92781  &   -1.907  &  203.5355  &  -29.8588  &            Injection 1\\
 33  &   4.0\times10^{-5} &  890.14676  &   -1.985  &  264.8075  &   30.4969  &            harmonic of a detector interference\\
 34  &   8.7\times10^{-5} &  912.66971  &    0.237  &    8.9448  &  -78.2029  &            harmonic of a detector interference\\
 35  &   8.7\times10^{-6} &  924.03645  &   -0.613  &  275.1312  &   51.0793  &            harmonic of a detector interference\\
 36  &   8.4\times10^{-5} &  952.61767  &   -0.479  &   50.3141  &  -73.5387  &            harmonic of a detector interference\\
 37  &   2.3\times10^{-4} &  992.81797  &   -0.514  &  281.6408  &   53.6023  &            harmonic of a detector interference\\
 38  &   1.2\times10^{-6} &  992.82278  &   -0.884  &   48.8682  &  -81.6560  &            harmonic of a detector interference\\
 39  &   1.7\times10^{-5} &  996.25027  &    0.239  &  271.2557  &   67.5053  &            Present only in H1
\\
 40  &   3.8\times10^{-7} &  996.25657  &   -1.660  &  111.3151  &  -76.5712  &            Present only in H1
\\
 41  &  3.8\times10^{-4} &  1000.81171  &   -1.333  &   92.7694  &  -85.8077  &            harmonic of a detector interference\\
 42  &  7.9\times10^{-5} &  1003.90928  &    0.242  &  274.4174  &   66.8185  &            Present only in H1
\\
 43  &  3.6\times10^{-7} &  1003.92034  &   -3.527  &  156.6592  &  -81.4408  &            Present only in H1
\\
 44  &  6.2\times10^{-4} &  1054.83208  &   -0.047  &  281.0917  &   46.6542  &            harmonic of a detector interference\\
 45  &  3.3\times10^{-4} &  1058.46127  &   -0.574  &   41.8203  &  -83.6738  &            harmonic of a detector interference\\
 46  &  2.7\times10^{-4} &  1142.02054  &   -1.289  &   18.8221  &  -85.2794  &            harmonic of a detector interference\\
 47  &  3.6\times10^{-4} &  1149.51676  &   -1.780  &  112.2596  &  -85.3719  &            harmonic of a detector interference\\
 48  &  4.0\times10^{-4} &  1163.07712  &   -0.461  &   71.0369  &  -77.8010  &            harmonic of a detector interference\\
 49  &  3.9\times10^{-4} &  1196.01380  &   -0.079  &   73.4466  &  -76.2717  &            harmonic of a detector interference\\
 50  &  3.6\times10^{-4} &  1201.09880  &   -0.391  &   75.8100  &  -76.7984  &            harmonic of a detector interference\\
 51  &  6.4\times10^{-4} &  1201.83843  &   -0.036  &   45.6877  &  -79.2300  &            harmonic of a detector interference\\
 52  &  3.0\times10^{-4} &  1210.30530  &    0.282  &   67.6575  &  -75.6512  &            harmonic of a detector interference\\
 53  &  0 & 1220.55246  &   -0.364  &  226.2481  &   -7.0719  &            Injection 7\\
 54  &  0 & 1220.55400  &   -0.078  &  218.8902  &  -32.1127  &            Injection 7\\
 55  &  8.3\times10^{-4} &  1224.35567  &   -1.593  &  269.1917  &   47.7573  &            harmonic of a detector interference\\
 56  &  4.7\times10^{-4} &  1250.03185  &   -0.632  &   58.5959  &  -81.8576  &            harmonic of a detector interference\\
 57  &  1.9\times10^{-4} &  1252.45409  &   -0.649  &   58.7640  &  -82.0590  &            harmonic of a detector interference\\
 58  &  2.3\times10^{-4} &  1253.19279  &   -0.946  &  276.3357  &   42.2750  &            harmonic of a detector interference\\
 59  &  6.8\times10^{-4} &  1287.31747  &   -0.692  &   81.6594  &  -77.4641  &            harmonic of a detector interference\\
 60  &  3.9\times10^{-4} &  1293.85609  &   -2.777  &  132.0928  &  -83.0378  &            harmonic of a detector interference\\
 61  &  6.5\times10^{-5} &  1310.08345  &   -2.338  &  113.5595  &  -82.3435  &            harmonic of a detector interference\\
 62  &  3.0\times10^{-4} &  1317.10722  &   -1.791  &  100.7620  &  -81.3209  &            harmonic of a detector interference\\
 63  &  1.1\times10^{-4} &  1381.05818  &   -0.239  &  279.5478  &   51.3271  &            harmonic of a detector interference\\
 64  &  4.3\times10^{-4} &  1383.22336  &   -2.009  &  108.8794  &  -81.2435  &            harmonic of a detector interference\\
 65  &  0 & 1393.54760  &   -2.011  &  323.8507  &    2.7705  &            Injection 4\\
 66  &  0 & 1393.55069  &   -1.496  &  336.9224  &  -25.1755  &            Injection 4\\
 67  &  4.0\times10^{-4} &  1411.31585  &   -2.444  &  115.2990  &  -80.9642  &            harmonic of a detector interference\\
 68  &  1.3\times10^{-4} &  1422.69979  &   -2.169  &  113.5662  &  -80.7357  &            harmonic of a detector interference\\
 69  &  8.3\times10^{-4} &  1468.11317  &    0.110  &  329.1319  &   -7.2537  &            Wandering frequency\\
 70  &  8.3\times10^{-4} &  1468.11329  &    0.115  &  332.3289  &  -15.9424  &            Wandering frequency\\
 71  &  2.4\times10^{-4} &  1573.10838  &   -0.908  &   78.8100  &  -79.5921  &            harmonic of a detector interference\\
 72  &  9.9\times10^{-4} &  1660.29579  &   -1.255  &  268.8632  &   52.7214  &            harmonic of a detector interference\\
 73  &  6.7\times10^{-4} &  1908.10543  &   -1.969  &  254.0502  &  -81.8658  &            Present in 1st 1/3 of the run
\\
 74  &  1.6\times10^{-6} &  1967.56836  &   -1.891  &  102.5901  &  -70.5698  &            Present only in L1
\\

\hline
\end{tabular}
\caption{{\td} pipeline outliers in the range of frequencies between 475 and 2000 Hz.
  The columns provide outliers false alarm probability (FAP) as well as the nominal frequencies and frequency derivatives, right ascensions and declinations
  found for the outliers, along with comments indicating the likely sources of the outliers.
Outliers described as "harmonics of a detector interference" are harmonics of an interference present in the detectors data
when no science data are taken.}
\label{tab:TDFstat_outliers}
\end{center}
\end{table*}

\end{document}